\documentclass[12pt,twoside,english]{article}
\usepackage[T1]{fontenc}
\usepackage[latin1]{inputenc}
\usepackage{geometry}
\usepackage{babel}
\usepackage{graphics}


\begin{document}
\def\A{\mathcal{A}}
\def\F{\mathcal{F}}

\begin{titlepage}

\setcounter{page}{1}
\rightline{}
\vfill
\begin{center}
 {\Large \bf An x-Space Analysis of Evolution Equations: Soffer's Inequality   
and the Non-forward Evolution}

\vfill
\vfill
 {\large Alessandro Cafarella,  
 Claudio Corianò and Marco Guzzi
}

\vspace{.12in}
 {\it Dipartimento di Fisica, Università di Lecce \\
 and INFN Sezione di Lecce \\ Via Arnesano 73100 Lecce, Italy \\
 and \\
alessandro.cafarella@le.infn.it, claudio.coriano@le.infn.it\\
 marco.guzzi@le.infn.it
}

\vspace{.075in}

\end{center}
\vfill

\begin{abstract}
We analize the use of algorithms based in x-space for the 
solution of renormalization group equations of DGLAP-type 
and test their consistency by studying bounds 
among partons distributions - in our specific case Soffer's inequality and the perturbative behaviour of the nucleon tensor charge -  
to next-to-leading order in QCD . A discussion 
of the perturbative resummation implicit in these expansions using Mellin moments is included. 
We also comment on the (kinetic) proof of positivity 
of the evolution of $h_1$, using a kinetic analogy and 
illustrate the extension of the algorithm to the evolution of generalized parton 
distributions. We prove positivity of the non-forward evolution in a special case 
and illustrate a Fokker-Planck approximation to it.

\end{abstract}
\smallskip

\end{titlepage}

\setcounter{footnote}{0}

\def\beq{\begin{equation}}
\def\eeq{\end{equation}}
\def\Q{{\bf Q}}
\def\z{{\zeta}}
\def\R{{\bf R}}
\def\beqn{\begin{eqnarray}}
\def\eeqn{\end{eqnarray}}
\def\ba{\begin{eqnarray}}
\def\ea{\end{eqnarray}}
\def\ie{{\it i.e.}}
\def\eg{{\it e.g.}}
\def\half{{\textstyle{1\over 2}}}
\def\nicefrac#1#2{\hbox{${#1\over #2}$}}
\def\third{{\textstyle {1\over3}}}
\def\quarter{{\textstyle {1\over4}}}
\def\m{{\tt -}}

\def\p{{\tt +}}

\def\slash#1{#1\hskip-6pt/\hskip6pt}
\def\slk{\slash{k}}
\def\GeV{\,{\rm GeV}}
\def\TeV{\,{\rm TeV}}
\def\y{\,{\rm y}}
\def\one {\bf  1}
\def\l{\langle}
\def\r{\rangle}

\setcounter{footnote}{0}
\newcommand{\beqa}{\begin{eqnarray}}
\newcommand{\eeqa}{\end{eqnarray}}
\newcommand{\eps}{\epsilon}

\pagestyle{plain}
\setcounter{page}{1}
\section{Introduction} 
One of the most fascinating aspects of the structure of the nucleon
is the study of the distribution of spin among its constituents, a topic of 
remarkable conceptual complexity which has gained a lot of attention 
in recent years. This study is entirely based on the classification and on the 
phenomenological modeling of all the leading-twist parton distributions, used 
as building blocks for further investigations in hadronic physics. 

There are various theoretical 
ways to gather information on these non-local matrix elements. One among the various 
possibilities is to discover sum rules connecting moments of these distributions to other 
fundamental observables. Another possibility is to discover 
bounds - or inequalities - among them and use these results in the process of their modeling. 
There are various bounds that can be studied, particularly in the context of the 
new generalized parton dynamics typical of the skewed distributions 
\cite{Ji,Radyushkin}. 
All these relations can be analized in perturbation theory and studied using 
the Renormalization Group (RG), although a complete description of their perturbative dynamics 
is still missing. This study, we believe, 
may require considerable theoretical effort since it involves a global understanding both of the (older) forward (DGLAP) dynamics and of the generalized 
new dynamics encoded in the skewed distributions. 

In this context, a program aimed at the study of various bounds in perturbation theory using primarily a parton dynamics in x-space has been outlined \cite{CafaCor}. 
This requires accurate algorithms to solve the equations up to 
next-to-leading order (NLO). Also, underlying this type of description is, in many cases, a 
probabilistic approach \cite{Teryaev1} which has some interesting consequences worth 
of a closer look . In fact, the DGLAP equation, viewed as a probabilistic process, 
can be rewritten in a {\em master form} which is at the root of some interesting 
formal developements. In particular, a wide set of results, available from the theory 
of stochastic processes, find their way in the study of the evolution. 
Two of us have elaborated on this issue in previous work \cite{CafaCor}
and proposed a Kramers-Moyal expansion of the DGLAP equation as an alternative 
way to describe its dynamics. Here, this analysis will be extended 
to the case of the non-forward evolution. 

With these objectives in mind, in this study we
test  x-space algorithms up to NLO and verify their accuracy using a stringent test: 
Soffer's inequality. As usual, we are bound to work with specific models of initial conditions.  
The implementations on which our analysis are based are general, 
with a varying flavour number $n_f$ at any threshold of intermediate quark mass 
in the evolution.  
Here, we address Soffer's inequality using an approach 
based on the notion of ``superdistributions'' 
\cite{Teryaev}, 
which are constructs designed to have a simple (positive) 
evolution thanks to the existence of an underlying master form \cite{Teryaev1,CafaCor}. 
The original motivation for using such a master form 
(also termed {\em kinetic} or {\em probabilistic}) 
to prove positivity has been presented in \cite{Teryaev}, 
while further extensions of these arguments have been presented in \cite{CafaCor}.  
In a final section we propose the extension of the evolution algorithm 
to the case of the skewed distributions, and illustrate its implementation 
in the non-singlet case. As for the forward case, 
numerical tests of the inequality are performed for 
two different models. We show that even starting from a saturated inequality at the 
lowest evolution scale, the various models differ significantly even for a moderate final 
factorization scale of $Q=100$ GeV. Finally, we illustrate in another application the 
evolution of the tensor charge and show that, in the models considered, differences 
in the prediction of the tensor charge are large.

\section{Prelude to x-space: A Simple Proof of Positivity of $h_1$ to NLO} 
There are some nice features of the parton dynamics, at least in the leading logarithmic approximation (LO), when viewed in x-space, once a suitable ``master form'' of the 
parton evolution equations is identified.    

The existence of such a master form, as firstly shown by Teryaev, 
is a special feature of the evolution equation 
itself. The topic has been addressed before 
in LO \cite{Teryaev} and reanalized in more detail in
\cite{CafaCor} where, starting from a kinetic interpretation of the 
evolution, a differential equation obtained 
from the Kramers-Moyal expansion of the DGLAP equation 
has also been proposed. 

The arguments of refs.~\cite{Teryaev,CafaCor} 
are built around a form of the evolution equation 
which has a simple kinetic interpretation and is written 
in terms of transition probabilities constructed from the kernels. 
  
The strategy used, at least in leading order,  
to demonstrate the positivity of the LO evolution for 
special combinations of parton distributions 
$\Q_\pm$ \cite{Teryaev}, to be defined below, or the NLO evolution for $h_1$, 
which we are going to address, is based on some results of
ref.\cite{Teryaev}, briefly reviewed here, in order 
to be self-contained. 

 A master equation is typically given by 
\beq
\frac{\partial }{\partial \tau}f(x,\tau)=\int dx'\left(
w(x|x') f(x',\tau) -w(x'|x) f(x,\tau)\right) dx'
\label{masterforms}
\eeq

and if through some manipulations, a DGLAP equation 

\beq
\frac{d q(x,Q^2)}{d \log( Q^2)} = \int_x^1 \frac{dy}{y} P(x/y)q(y,Q^2),
\eeq
with kernels $P(x)$, is rewritten in such a way to resemble  
eq. (\ref{masterforms})  

\beq
\frac{d}{d \tau}q(x,\tau) = \int_x^1 dy \hat{P}\left(\frac{x}{y}\right)\frac{q(y,\tau)}{y}
-\int_0^x \frac{dy}{y}\hat{ P}\left(\frac{y}{x}\right)\frac{q(x,\tau)}{x},
\label{bolz}
\eeq
with a (positive) transition probability 
\beq
w(x|y)= \frac{\alpha_s}{2 \pi} \hat{P}(x/y)\frac{\theta(y > x)}{y}
\eeq
then positivity of the evolution is established. 

For equations of non-singlet type, such 
as those evolving $q^{(-)}=q - \bar{q}$, the valence quark distribution, 
or $h_1$, the transverse spin distribution, 
this rewriting of the equation is possible, at least in LO. 
NLO proofs are, in general, impossible to construct by this method, 
since the kernels turn out, in many cases, to be negative. The only possible proof, in these cases, is just a numerical one, for suitable (positive) 
boundary conditions observed by the initial form of the parton distributions. 
Positivity of the evolution is then a result of an unobvious interplay between 
the various contributions to the kernels in various regions in x-space.  

In order to discuss the probabilistic version of the DGLAP equation it 
is convenient to separate the bulk contributions of the kernels $(x<1)$ from the 
edge point contributions at $x=1$. For this purpose 
we recall that the structure of the kernels is, in general, given by 
\beq
P(z) = \hat{P}(z) - \delta(1-z) \int_0^1 \hat{P}(z)\, dz,
\label{form}
 \eeq
where the bulk contributions $(z<1)$ and the edge point contributions 
$(\sim \delta(z-1))$ have been explicitely separated.
We focus on the transverse spin distributions as an example. 
With these prerequisites, 
proving the LO and NLO positivity of the transverse spin distributions 
is quite straightforward, but requires a numerical inspection of the transverse 
kernels. Since the evolutions for $\Delta_T q^{(\pm)}\equiv h^q_1$ are purely non-singlet, 
diagonality in flavour of the subtraction terms $(\sim \int_0^x w(y|x)q(x,\tau))$ 
is satisfied, while the edge-point subtractions can be tested 
to be positive numerically. 
We illustrate the explicit construction of the master equation for $h_1$ in LO, since extensions to NLO of this construction are rather straighforward. 

In this case the LO kernel is given by 

\beqn
\Delta_{T}P^{(0)}_{qq}(x)= C_{F}\left[\frac{2}{(1-x)_{+}}-2 +\frac{3}{2}\delta(1-x)\right] 
\eeqn
and by some simple manipulations we 
can rewrite the corresponding evolution equation 
in a suitable master form. That this is possible is an elementary fact 
since the subtraction terms 
can be written as integrals of a positive function. For instance, 
a possibility is 
to choose the transition probabilities
\beqa
w_1[x|y] &=& \frac{C_F}{y}\left(\frac{2}{1- x/y} - 2 \right)
\theta(y>x) \theta(y<1)\nonumber \\
w_2[y|x] &=& \frac{C_F}{x} \left(\frac{2}{1- y/x} - \frac{3}{2}\right)
\theta(y > -x)\theta(y<0)
\nonumber \\
\eeqa
which reproduce the evolution equation for $h_1$ in master form

\beq
\frac{d h_1}{d \tau}= \int_0^1 dy w_1(x|y)h_1(y,\tau) 
-\int_0^1 dy w_2(y|x) h_1(x,\tau).
\label{masterix}
\eeq

\begin{figure}
{\centering \resizebox*{12cm}{!}{\rotatebox{-90}{\includegraphics{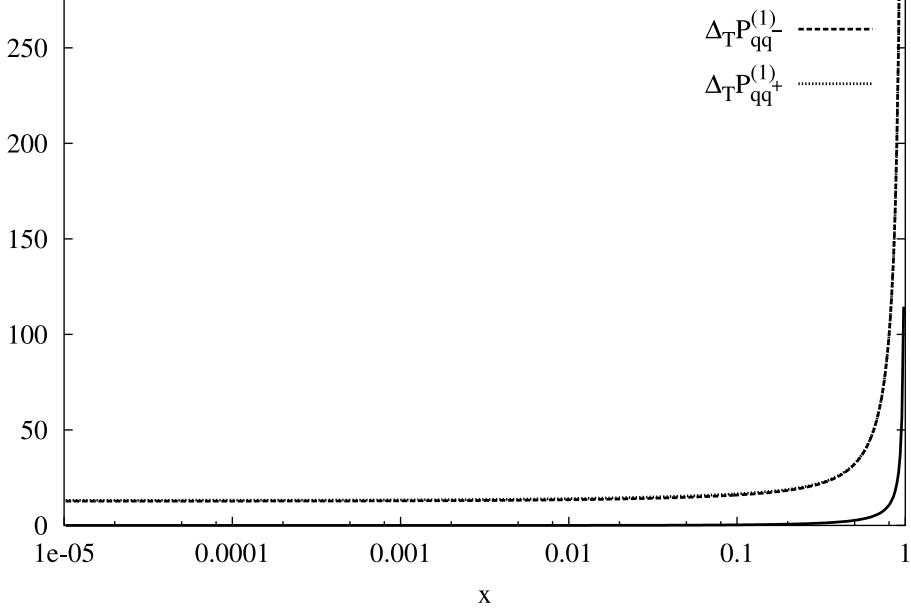}}} \par}
\caption{Plot of the transverse kernels.}
\label{transversekernels}
\end{figure}

The NLO proof of positivity is also rather straightforward. 
For this purpose we have analized numerically the behaviour of the NLO kernels both 
in their bulk region and at the edge-point. 
We show in Table 1 of Appendix B results 
for the edge point contributions to NLO for both of 
the $\Delta_T P^{(1)}_\pm$ components, 
which are numerically the same.
There we have organized these terms in the form $\sim C\delta(1-x)$ with
\beq
C=-\log(1- \Lambda) A + B 
\label{sub}, 
\eeq
with A and B being numerical coefficients depending on the number 
of flavours included in the kernels. 
The (diverging) logarithmic contribution ($\sim \int_0^\Lambda dz/(1-z)$) 
have been regulated by a cutoff. This divergence 
in the convolution cancels when these terms are combined with the divergence at 
$x=1$ of the first term of the master equation (\ref{masterix}) 
for all the relevant components 
containing ``+'' distributions. As for the 
bulk contributions $(x<1)$, positivity up to NLO of the transverse kernels 
is shown numerically in Fig. (\ref{transversekernels}). 
All the conditions of positivity are therefore satisfied and therefore 
the $\Delta_{T\pm}q$ distributions evolve positively up to NLO. 
The existence of a master form of the equation is then guaranteed.

Notice that the NLO positivity of $\Delta_{T\pm}q$ implies positivity of the
nucleon tensor charge \cite{JJ}
\beq
\delta q\equiv\int_0^1 dx \left( h_1^q(x) - h_1^{\bar{q}(x)}\right)
\eeq
for each separate flavour for positive initial conditions. 
As we have just shown, this proof of positivity is very short, as far as one 
can check numerically that both components of eq.(\ref{masterix}) 
are positive. 

\section{Soffer's inequality}
Numerical tests of Soffer's inequality can be performed 
either in moment space or, as we are going to illustrate 
in the next section, directly in x-space, using suitable 
algorithms to capture the perturbative nature of the evolution.  
We recall that Soffer's inequality
\beq
|h_1(x)| < q^+(x)
\eeq
sets a bound on the transverse spin distribution $h_1(x)$ in terms of the
components of the positive helicity component of the quarks, for a given flavour.
An original proof of Soffer's inequality 
in LO has been discussed in ref.\cite{Barone}, while 
in \cite{Teryaev} an alternative proof was presented, based 
on a kinetic interpretation of the evolution equations. 

We recall that $h_1$, also denoted by the symbol 
\begin{equation}
\Delta _{T}q(x,Q^{2})\equiv q^{\uparrow }(x,Q^{2})-q^{\downarrow }(x,Q^{2}),
\end{equation} 
has the property
of being purely non-singlet and of appearing at leading twist. It is
identifiable in transversely polarized 
hadron-hadron collisions and not in Deep Inelastic Scattering (from now on we will 
omit sometime the x-dependence in the kernels and in the distributions when obvious).
In the following we will use interchangeably the notations $h_1\equiv h_1^q$ 
and $\Delta_T q$ to 
denote the transverse asymmetries. We introduce also the combinations 
\beqa
\Delta_T(q + \bar{q}) &=& h_1^q + h_1^{\bar{q}} \nonumber \\
\Delta_T q^{(-)}=\Delta_T(q - \bar{q}) &=& h_1^q - h_1^{\bar{q}} \nonumber \\
\Delta_T q^{(+)} &=& \sum_i \Delta_T(q_i + \bar{q}_i) \nonumber \\
\eeqa
where we sum over the flavor index $(i)$, and we have introduced singlet and non-singlet 
contributions for distributions of fixed helicities 
\beqa
q_+^{(+)}&=&\sum_i\left( q_{+ i} + \bar{q}_{+ i}\right)\nonumber \\
q_+^{(-)}&=& q_{+ i} -\bar{q}_{+ i}\equiv \Sigma. \nonumber \\
\eeqa
In our analysis we solve all the equations in the helicity basis and reconstruct 
the various helicities after separating singlet and non-singlet sectors. 
We mention that 
the non-singlet sector is now given by a set of 2 equations, each involving 
$\pm$ helicities and the singlet sector is given by a 4-by-4 matrix.   

In the singlet sector we have 

\begin{eqnarray}
{dq_+^{(+)} \over{dt}}=
{\alpha_s \over {2 \pi}} (P_{++}^{qq}\otimes q_+^{(+)}+
P_{+-}^{qq} \otimes q_-^{(-)}  \nonumber \\
+P_{++}^{qG} \otimes G_++
P_{+-}^{qG} \otimes G_-),
\nonumber \\
{dq_-^{(+)}(x) \over{dt}}=
{\alpha_s \over {2 \pi}} (P_{+-} \otimes q_+^{(+)} +
P_{++}  \otimes q_-^{(+)} \nonumber \\
+P_{+-}^{qG} \otimes G_+ +
P_{++}^{qG} \otimes G_-),  \nonumber \\
{dG_+(x) \over{dt}}=
{\alpha_s \over {2 \pi}} (P_{++}^{Gq} \otimes q_+^{(+)}+
P_{+-}^{Gq} \otimes q_-^{(+)} \nonumber \\
+P_{++}^{GG}\otimes G_+ +
P_{+-}^{GG} \otimes G_-),  \nonumber \\
{dG_-(x) \over{dt}}=
{\alpha_s \over {2 \pi}} (P_{+-}^{Gq} \otimes q_+^{(+)} +
P_{++}^{Gq} \otimes q_-^{(+)} \nonumber \\
+P_{+-}^{GG} \otimes G_+ +
P_{++}^{GG} \otimes G_-).
\label{hs}\end{eqnarray}

while the non-singlet (valence) analogue of this equation is also easy to
write down
\begin{eqnarray}
{dq_{+ i}^{(-)}(x) \over{dt}}=
{\alpha_s \over {2 \pi}} (P^{NS}_{++} \otimes q_{+ i}^{(-)}+
P^{NS}_{+-} \otimes q_{-}^{(-)}(y)), \nonumber \\
{dq_{- i}^{(-)}(x) \over{dt}}=
{\alpha_s \over {2 \pi}} (P^{NS}_{+-} \otimes q_{+}^{(-)}+
P^{NS}_{++} \otimes q_{- i}^{(-)}).
\label{h}\end{eqnarray}
Above, $i$ is the flavor index, $(\pm)$ indicate $q\pm \bar{q}$ components and the lower subsctipt $\pm$ stands for the helicity.

Similarly to the unpolarized case the flavour reconstruction is done by adding 
two additional equations for each flavour in the helicity $\pm$
\beq
\chi_{\pm,i}= q_{\pm i}^{(+)}- \frac{1}{n_f}q^{(+)}_\pm
\eeq
whose evolution is given by 
\beqa
{d \chi_{+ i}^{(-)}(x) \over{dt}} &=&
{\alpha_s \over {2 \pi}} (P^{NS}_{++} \otimes \chi_{+ i} +
P^{NS}_{+-} \otimes \chi_{- i}), \nonumber \\
{d \chi_{- i} (x) \over{dt}} &=&
{\alpha_s \over {2 \pi}} (P^{NS}_{+-} \otimes \chi_{+ i} +
P^{NS}_{++} \otimes \chi_{- i}). \nonumber \\
\label{h11}
\end{eqnarray}

The reconstruction of the various contributions in flavour space 
for the two helicities is finally done 
using the linear combinations 
\beq
q_{\pm i}=\frac{1}{2}\left( q_{\pm i}^{(-)} + \chi_{\pm i} +\frac{1}{n_f}q_{\pm}^{(+)}\right).
\eeq

We will be needing these equations below when we present
a proof of positivity up to LO, and we will thereafter proceed with a NLO implementation of these and other evolution equations. For this we will be needing some more notations. 

We recall that the following relations are also true to all orders 
\beqa
P(x) &=&\frac{1}{2}\left( P_{++}(x) + P_{+-}(x)\right)\nonumber \\
&=&\frac{1}{2}\left( P_{--}(x) + P_{-+}(x)\right)\nonumber 
\eeqa
between polarized and unpolarized $(P)$ kernels 
and 
\beq
P_{++}(x) =  P_{--}(x),\,\,\,P_{-+}(x)=P_{+-}(x)
\eeq
relating unpolarized kernels to longitudinally polarized ones. 
Generically, the kernels of various type are expanded up to NLO as 
\beq
P(x)= \frac{\alpha_s}{2 \pi} P^{(0)}(x) + \left(\frac{\alpha_s}{2 \pi}\right)^2 P^{(1)}(x),
\eeq
and specifically, in the transverse case we have

\begin{eqnarray} \label{pm}
\Delta_T P_{qq,\pm}^{(1)} &\equiv& \Delta_T P_{qq}^{(1)} \pm \Delta_T 
P_{q\bar{q}}^{(1)} \; , \\
\end{eqnarray}
with the corresponding evolution equations 

\begin{equation} \label{evol3}
\frac{d}{d\ln Q^2} \Delta_T q_{\pm} (Q^2) = \Delta_T P_{qq,\pm} 
(\alpha_s (Q^2))\otimes \Delta_T q_\pm (Q^2) \; .
\end{equation}

We also recall that the kernels in the helcity basis in LO are given by 
\beqa
P_{NS\pm,++}^{(0)} &=&P_{qq, ++}^{(0)}=P_{qq}^{(0)}\nonumber \\
P_{qq,+-}^{(0)}&=&P_{qq,-+}^{(0)}= 0\nonumber \\
P_{qg,++}^{(0)}&=& n_f x^2\nonumber \\
P_{qg,+-}&=& P_{qg,-+}= n_f(x-1)^2 \nonumber \\
P_{gq,++}&=& P_{gq,--}=C_F\frac{1}{x}\nonumber \\ 
P_{gg,++}^{(0)}&=&P_{gg,++}^{(0)}= N_c
\left(\frac{2}{(1-x)_+} +\frac{1}{x} -1 -x - x^2 \right) +{\beta_0}\delta(1-x) \nonumber \\
P_{gg,+-}^{(0)}&=& N_c
\left( 3 x +\frac{1}{x} -3 - x^2 \right). 
\label{stand1}
\eeqa

An inequality, such as Soffer's inequality, can be stated as positivity condition 
for suitable linear combinations of parton distributions \cite{Teryaev} 
and this condition can be analized - as we have just shown 
for the $h_1$ case -  in a most direct way using the master form.

For this purpose consider the linear valence combinations
\beqa
\Q_+ &=& q_+ + h_1 \nonumber \\
\Q_-  &=& q_+ - h_1 \nonumber \\
\eeqa
which are termed ``superdistributions'' in ref.\cite{Teryaev}. Notice that a proof 
of positivity of the $\Q$ distributions is equivalent to verify Soffer's inequality. 
However, given the mixing of singlet and non-singlet sectors, the analysis of 
the master form is, in this case, more complex. As we have just mentioned, what can spoil the proof of 
positivity, in general, is the negativity of the kernels to higher order. 
We anticipate here the result that we will illustrate below where we show 
that a LO proof of the positivity of the evolution for $\Q$ can be established using 
kinetic arguments, being the kernels are positive at this order. However 
we find 
that the NLO kernels do not satisfy 
this condition. 
In any case, let's see how the identification of such master form proceeds in general. 
We find useful to illustrate the result using the separation between singlet and non-singlet 
sectors. In this case we introduce the combinations

\beqa
\Q_\pm^{(-)} &=& q_+^{(-)} \pm h_1^{(-)} \nonumber \\
\Q_\pm^{(+)} &=& q_+^{(+)} \pm h_1^{(+)} \nonumber \\
\label{separation}
\eeqa
with $h_1^{(\pm)}\equiv \Delta_T q^{(\pm)}$.  

Differentiating these two linear combinations  (\ref{separation}) we get
\beqa
\frac{d \Q_\pm^{(-)}}{d\log(Q^2)}= P^{NS}_{++} q_+^{(-)} 
+ P^{NS}_{+ -} q_-^{(-)} \pm P_T h_1^{(-)} \nonumber \\
\eeqa
which can be rewritten as
\beqa
\frac{d \Q_+^{(-)}}{d\log(Q^2)} &=& \frac{1}{2}\left(P_{++}^{(-)} + P_T^{(-)}\right)\Q_+^{(-)}
 + \frac{1}{2}\left(P_{++}^{(-)} - P_T^{(-)}\right)\Q_-^{(-)} + P_{+ -}^{(-)}q_-^{(-)} \nonumber \\
\frac{d \Q_+^{(-)}}{d\log(Q^2)} &=& \frac{1}{2}\left(P_{++}^{(-)} - P_T^{(-)}\right)\Q_+^{(-)}
+ \frac{1}{2}\left(P_{++}^{(-)} + P_T\right)^{(-)}\Q_-^{(-)} + P_{+ -}^{(-)}q_-^{(-)} \nonumber \\
\eeqa
with $P^{(-)}\equiv P^{NS}$ being the non-singlet (NS) kernel. 

At this point we define the linear combinations
\beqa
{\bar{P}^Q}_{+\pm}= \frac{1}{2}\left(P_{++} \pm P_T\right)
\eeqa
and rewrite the equations above as
\beqa
\frac{d \Q_+ i}{d\log(Q^2)} &=& \bar{P}^Q_{ ++}\Q_{i+}
 + \bar{P}^Q_{+-}\Q_{i-} + P^{qq}_{+ -}q_{i-} \nonumber \\
\frac{d \Q_{i+}}{d\log(Q^2)} &=& \bar{P}^Q_{+-}\Q_{i+}
 + \bar{P}^Q_{++}\Q_{i-} + P_{+ -}^{qq}q_{i-} \nonumber \\
\label{pos1}
\eeqa
where we have reintroduced $i$ as a flavour index.
From this form of the equations it is easy to establish the leading order positivity of the evolution, after checking the positivity of the kernel and the existence of a master form.

The second non-singlet sector is defined via the variables 

\beq
\chi_{i\pm}=q_{i \pm}^{(+)} - \frac{1}{n_f}q_{i\pm}^{(+)}
\eeq
which evolve as non-singlets 
and the two additional distributions 
\beqa
\Q_{\chi i,\pm}= \chi_{i +} \pm h_1^{i(+)}.
\eeqa
Also in this case we introduce  the kernels 
\beqa
{\bar{P}^{Q_\chi}}_{+\pm} &=& \frac{1}{2}\left(P_{++} \pm 
\Delta_T P^{(+)}\right)
\eeqa
to obtain the evolutions 
\beqa
\frac{d \Q_{\chi i+}}{d\log(Q^2)} &=& \bar{P}^{Q_\chi}_{ ++}\Q_{\chi i+}
 + \bar{P}^{Q_\chi}_{+-}\Q_{\chi i-} + P^{qq}_{+ -}\chi_{i-} \nonumber \\
\frac{d \Q_{i+}}{d\log(Q^2)} &=& \bar{P}^Q_{\chi +-}\Q_{\chi i+}
 + \bar{P}^{Q_\chi}_{++}\Q_{\chi i-} + P_{+ -}^{qq}\chi_{i-}. \nonumber \\
\label{pos12}
\eeqa

For the singlet sector, we simply define $Q_+^{(+)}=q^{(+)}$, and the 
corresponding evolution is similar to the singlet equation of the helicity basis. 
Using the equations above, 
the distributions $\Q_{i\pm}$ are then reconstructed as 
\beq
\Q_{i\pm} = \frac{1}{2}\left(\Q_{i \pm}^{(-)} + 
\Q_{\chi i \pm}^{(-)} + \frac{1}{n_f}Q_+^{(+)}\right)
\eeq
and result positive for any flavour if the addends are positive as well. 
However, as we have just mentioned, positivity of all the kernels introduced above is easy to check numerically to LO, together with their diagonality in flavour which guarantees the existence 
of a master form.  

As an example, consider the LO evolution of $\Q\pm$. 
The proof of positivity is a simple consequence of the structure of eq.~(\ref{pos1}).
In fact the edge-point
contributions appear only in $P^Q_{++}$, i.e. they are diagonal in the evolution of
$\Q_{\pm}$. The inhomogenous terms on the right hand side of (\ref{pos1}), proportional to
$q_-$ are are harmless, since the $P_{+-}$ kernel has no edge-point contributions. Therefore under 1) diagonality in flavour of the subtraction terms and 
2) positivity of first and second term 
(transition probabilities) we can have positivity of the evolution. A 
refined arguments to support this claim has been presented in \cite{CafaCor}.    

\begin{figure}
{\centering \resizebox*{12cm}{!}{\rotatebox{-90}{\includegraphics{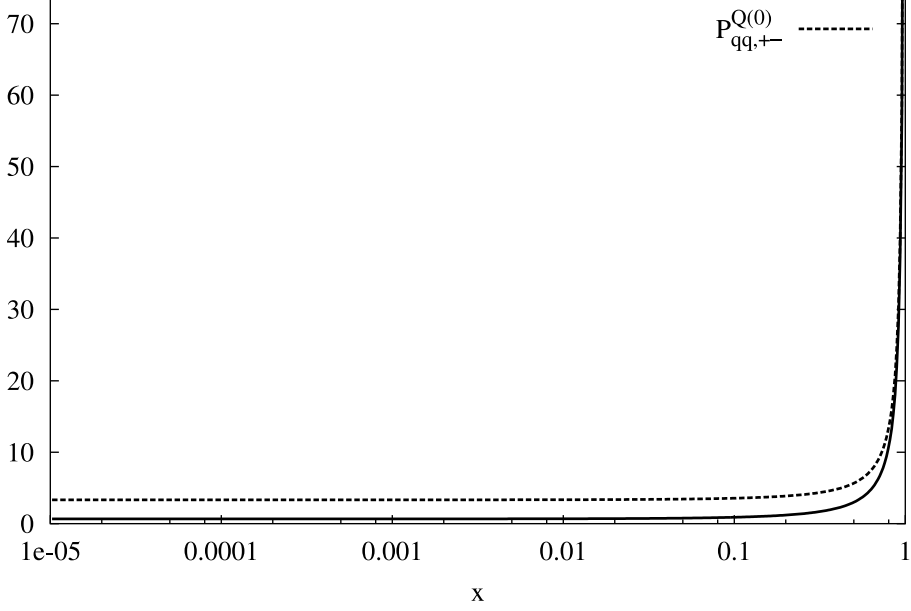}}} \par}
\caption{Plot of the LO kernels for the $\Q$ distributions}
\label{Qkernels}
\end{figure}

\begin{figure}
{\centering \resizebox*{12cm}{!}{\rotatebox{-90}{\includegraphics{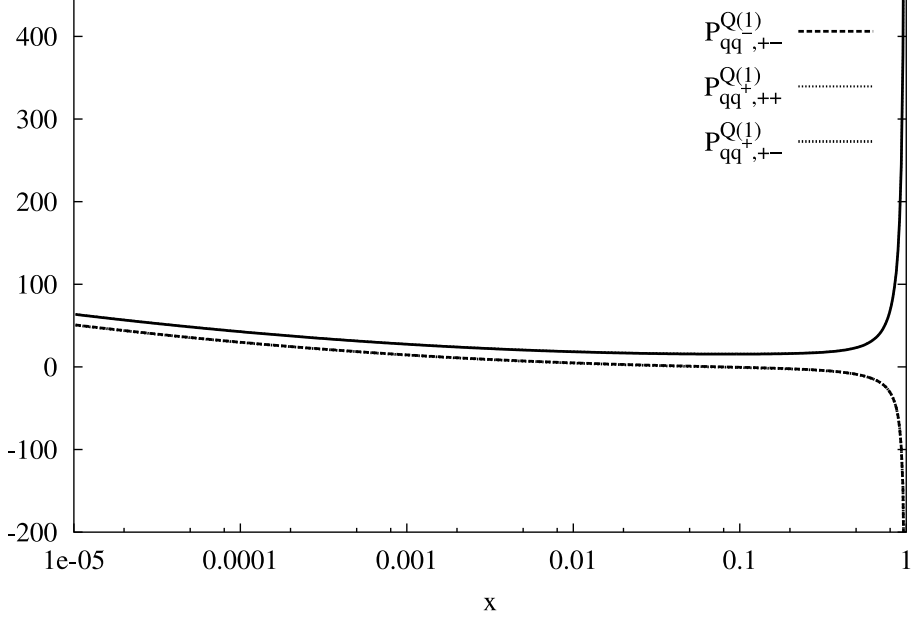}}} \par}
\caption{Plot of the NLO kernels for the $\Q$ distributions, 
showing a negative behaviour at large $x$}
\label{QNLOkernels}
\end{figure}

This construction is not valid to NLO. In fact, while the features of flavour 
diagonality of the 
master equation  are satisfied, the transition probabilities $w(x,y)$ 
are not positive in the whole $x,y$ range. The existence 
of a crossing from positive to negative values in $P^{\Q}_{++}$ 
can, in fact, be established quite easily using a numerical analysis. We illustrate in Figs. 
(\ref{Qkernels}) and (\ref{QNLOkernels}) plots of the $\Q$ kernels at LO and NLO, 
showing that, at NLO, the requirement of positivity of some components is violated.  
The limitations of this sort of proofs -based on kinetic arguments- are strictly 
linked to the positivity of the transition probabilities once a master form of the 
equation is identified. 
\section{ An x-space Expansion}
We have seen that NLO proofs of positivity, can be -at least partially- obtained 
only for suitable sets of boundary conditions. To this purpose, we choose to 
investigate the numerical behaviour of the solution using x-space 
based algorithms which need to be tested up to NLO.   

Our study validates a method which can be used to solve evolution equations 
with accuracy in leading and in next-to-leading order. The method is entirely 
based on an expansion \cite{Rossi} used in the context of spin physics \cite{Gordon}
and in supersymmetry \cite{Coriano}. An interesting feature of the
expansion, once combined with Soffer's inequality, is to generate an infinite set of 
relations among the scale invariant coefficients $(A_n, B_n)$ which characterize it. 

In this approach, the NLO expansion of the distributions in the DGLAP equation is generically given by
\beq
f(x,Q^2)=\sum_{n=0}^{\infty} \frac{A_n(x)}{n!}\log^n
\left(\frac{\alpha(Q^2)}{\alpha(Q_0^2)}\right) +
\alpha(Q^2)\sum_{n=0}^\infty \frac{B_n(x)}{n!}\log^n
\left(\frac{\alpha(Q^2)}{\alpha(Q_0^2)}\right)
\label{expansion}
 \eeq
where, to simplify the notation, we assume a short-hand matrix notation for all the convolution products.
Therefore $f(x,Q^2)$ stands for a vector having as components 
any of the helicities of the various flavours  $(\Q_{\pm},q_\pm,G_\pm)$.
The ansatz implies a tower of recursion relations once the running coupling
is kept into account and implies that 
\beq
A_{n+1}(x) =  -\frac{2}{\beta_0}P^{(0)}\otimes A_n(x)
\label{recur}
\eeq
to leading order
and
\beqa
B_{n+1}(x) & = & - B_n(x)- \left(\frac{\beta_1}{4 \beta_0} A_{n+1}(x)\right)
- \frac{1}{4 \pi\beta_0}P^{(1)}\otimes A_n(x) -\frac{2}{\beta_0}P^{(0)}
\otimes B_n(x) \nonumber \\
 & = &  - B_n(x) + \left(\frac{\beta_1}{2 \beta_0^2}P^{(0)}\otimes A_n(x)\right)
 \nonumber \\
&&- \frac{1}{4 \pi\beta_0}P^{(1)}\otimes A_n(x) -\frac{2}{\beta_0}P^{(0)}
\otimes B_n(x),  \nonumber \\
\label{recur1}
\eeqa
to NLO, relations which are solved with the initial condition $B_0(x)=0$.
The initial conditions for the coefficients $ A_0(x)$ and $B_0(x)$ are specified 
with $q(x,Q_0^2)$ as a leading order ansatz for the initial
distribution
\beq
A_0(x)= \delta(1-x)\otimes q(x,Q_0^2)\equiv q_0(x)
\eeq
which also requires $B_0(x)=0$, since we have to
satisfy the boundary condition 
\beq
A_0(x) + \alpha_0 B_0(x)= q_0(x).
\label{bdry}
\eeq

Again, other boundary choices are possible for $A_0(x)$ and $B_0(x)$
as far as (\ref{bdry}) is fullfilled.

If we introduce Rossi's expansion for $h_1$, $q_+$, and the linear combinations $\Q_\pm$ (in short form)

\beqa
h_1 &\sim&\left(A^{h}_n,B^{h +}_n\right)\nonumber \\
q_\pm &\sim&\left(A^{q_\pm}_n,B^{q_\pm}_n\right)\nonumber \\
\Q_\pm &\sim&\left(A^{Q +}_n,B^{Q +}_n\right) \nonumber \\
\eeqa

we easily get the inequalities
\beq
(-1)^n\left(A^{q_+}_n + A^{h}_n\right) >0
\eeq
and
\beq
(-1)^n\left(A^{q_+}_n - A^{h}_n\right) >0
\eeq
valid to leading order,which we can check numerically. 
Notice that the signature factor has to be included due to the alternation
in sign of the expansion.
To next to leading order we obtain 

\beq
 (-1)^{n+1} \left(A^{q_+}_n(x) +\alpha(Q^2) B^{q_+}_n (x)\right)\, 
< (-1)^n \left(A^{h}_n(x)\,+\alpha(Q^2) B^{h}_n (x)\right)\,  <\, (-1)^n \left(A^{q_+}_n(x) +\alpha(Q^2) 
B^{q_+}_n (x)\right), 
\eeq
valid for $n\geq 1$, obtained after identification of the corresponding 
logarithmic powers $\log\left(\alpha(Q^2)\right)$ at any $Q$. 
In general, one can assume a saturation of the inequality at the initial evolution scale 
\beq
\Q_-(x,Q_0^2)=h_1(x,Q_0^2) -\frac{1}{2}q_+(x,Q_0^2)=0. 
\eeq
This initial condition has been evolved in $Q$ solving the equations 
for the $\Q_\pm$ distributions  to NLO.

\section{Relations among moments}
In this section we elaborate on the relation between the coefficients of the 
recursive expansion as defined above and the standard solution 
of the evolution equations in the space of Mellin moments. We will show that the 
two solutions can be related in an unobvious way. 

Of our concern here is the relation between 
the Mellin moments of the coefficients appearing in the expansion 
\beqa
A(N) &=& \int_0^1\,dx \, x^{N-1} A(x)\nonumber \\
B(N) &=&\int_0^1\,dx \, x^{N-1} B(x) \nonumber \\
\eeqa
and those of the distributions  
\beq
\Delta_T q_{\pm} (N,Q^2)=\int_0^1\,dx \, x^{N-1} \Delta_T(x,Q^2)).
\eeq 
For this purpose we recall that the general (non-singlet) solution to NLO for the latter moments is given by 
\begin{eqnarray} \label{evsol}
\nonumber
\Delta_T q_{\pm} (N,Q^2) &=& K(Q_0^2,Q^2,N)
\left( \frac{\alpha_s (Q^2)}{\alpha_s (Q_0^2)}\right)^{-2\Delta_T 
P_{qq}^{(0)}(N)/ \beta_0}\! \Delta_T q_{\pm}(N, Q_0^2)
\label{solution}
\end{eqnarray}
with the input distributions $\Delta_T q_{\pm}^n (Q_0^2)$ at the input scale 
$Q_0$
and where we have set 
\beq
K(Q_0^2,Q^2,N)= 1+\frac{\alpha_s (Q_0^2)-
\alpha_s (Q^2)}{\pi\beta_0}\!
\left[ \Delta_T P_{qq,\pm}^{(1)}(N)-\frac{\beta_1}{2\beta_0} \Delta_T 
P_{qq\pm}^{(0)}(N) \right]. 
\eeq
In the expressions above we have introduced the corresponding moments for the LO and NLO kernels 
($\Delta_T P_{qq}^{(0),N}$,
$ \Delta_T P_{qq,\pm}^{(1),N})$. 

We can easily get the relation between the moments of the coefficients of the non-singlet
x-space expansion and those of the parton distributions at any $Q$, as expressed by eq.~(\ref{solution})
\beq
A_n(N) + \alpha_s B_n(N)=\Delta_T q_\pm(N,Q_0^2)K(Q_0,Q,N)\left(\frac{-2 \Delta_T P_{qq}(N)}{\beta_0}\right)^n.
\label{relation}
\eeq

As a check of this expression, notice that the initial condition is easily obtained from  
(\ref{relation}) setting $Q\to Q_0, n\to 0$, thereby obtaining 
\beq
A_0^{NS}(N) + \alpha_s B_0^{NS} (N)= \Delta_T q_\pm(N,Q_0^2)
\eeq
which can be solved with $A_0^{NS}(N)=\Delta_T q_\pm(N,Q_0^2)$ and 
$B_0^{NS} (N)=0$. 

It is then evident that the expansion (\ref{expansion}) involves a resummation of the logarithmic contributions, as shown in eq.~(\ref{relation}). 

In the singlet sector we can work out a similar relation both to LO 

\beq
A_n(N) = e_1\left(\frac{-2 \lambda_1}{\beta_0}\right)^n 
+e_2 \left(\frac{-2 \lambda_2}{\beta_0}\right)^n 
\eeq

with 
\beqa
e_1 &=& \frac{1}{\lambda_1 - \lambda_2}\left( P^{(0)}(N)- \lambda_2 \one \right)
\nonumber \\
e_2 &=& \frac{1}{\lambda_2 - \lambda_1}\left( - P^{(0)}(N) + \lambda_1 \one\right)
\nonumber \\
\lambda_{1,2}&=& \frac{1}{2}\left( 
P^{(0)}_{qq}(N) + P^{(0)}_{gg}(N) \pm \sqrt{\left(P^{(0)}_{qq}(N)- P^{(0)}_{gg}(N)\right)^2  
+ 4 P^{(0)}_{qg}(N)P^{(0)}_{gq}(N)}\right),
\eeqa
and to NLO 
\beq
A_n(N) + \alpha_s B_n(N) = \chi_1\left(\frac{-2 \lambda_1}{\beta_0}\right)^n 
+\chi_2 \left(\frac{-2 \lambda_2}{\beta_0}\right)^n 
\eeq

where 
\beqa
\chi_1 &=& e_1 + \frac{\alpha}{2 \pi}\left( \frac{-2}{\beta_0}e_1 \R e_1 
+\frac{ e_2 \R e_1}{\lambda_1 - \lambda_2 - \beta_0/2}\right) \nonumber \\
\chi_2 &=& e_2 + \frac{\alpha}{2 \pi}\left( \frac{-2}{\beta_0}e_2 \R e_2  
+\frac{ e_1 \R e_2}{\lambda_2 - \lambda_1 - \beta_0/2}\right)\nonumber \\
\eeqa
with 
\beq
\R= P^{(1)}(N) -\frac{\beta_1}{2 \beta_0}P^{(0)}(N).
\eeq
Notice that $A_n(N)$ and $B_n(N)$, $P^{(0)}(N)$, $P^{(1)}(N)$, in this case, 
are all 2-by-2 singlet matrices. Prior to discuss some phenomenological 
application of this method, 
left to a final section,  we will now illustrate its application 
to the case of the evolution of $h_1$ up to NLO.
Its extension to the case of the generalized parton distributions 
will follow afterwards.

\section{An Example: The Evolution of the Transverse Spin Distributions}
LO and NLO recursion relations for the coefficients of the expansion 
can be worked out quite easily, although the numerical implementation of these 
equations is far from being obvious. Things are somehow 
simpler to illustrate in the case of simple non-singlet evolutions, such 
as those involving transverse spin distributions, as we are going to show below. 
Some details and definitions can be found in the appendix.

For the first recursion relation (eq. (\ref{recur})) we have
\beqn
&&A^{\pm}_{n+1}(x)=-\frac{2}{\beta_{0}}\Delta_{T}P^{(0)}_{qq}(x)\otimes A^{\pm}_{n}(x)=\nonumber\\ 
&&C_{F}\left(-\frac{4}{\beta_{0}}\right)\left[\int^{1}_{x}\frac{dy}{y}\frac{y A^{\pm}_{n}(y) - x A^{\pm}_{n}(x)}{y-x} + A^{\pm}_{n}(x) \log(1-x)\right]+\nonumber\\
&&C_{F}\left(\frac{4}{\beta_{0}}\right) \left(\int_{x}^{1}\frac{dy}{y} A^{\pm}_{n}(y)\right) + C_{F}\left(-\frac{2}{\beta_{0}}\right)\frac{3}{2} A^{\pm}_{n}(x)\,.
\eeqn
As we move to NLO, it is convenient to summarize 
the structure of the transverse kernel $\Delta_{T}P^{\pm, (1)}_{qq}(x)$ as  

\beqn
&&\Delta_{T}P^{\pm, (1)}_{qq}(x)= K^{\pm}_{1}(x)\delta(1-x) + K^{\pm}_{2}(x)S_{2}(x) +K^{\pm}_{3}(x)\log(x)\nonumber\\
&&+ K^{\pm}_{4}(x)\log^{2}(x) +K^{\pm}_{5}(x)\log(x)\log(1-x) + K^{\pm}_{6}(x)\frac{1}{(1-x)_{+}} + K^{\pm}_{7}(x)\,.    
\eeqn

Hence, for the $(+)$ case we have 

\beqn
&&\Delta_{T}P^{+, (1)}_{qq}(x)\otimes A^{+}_{n}(x) = K^{+}_{1} A^{+}_{n}(x) + \int^{1}_{x}\frac{dy}{y}\left[K^{+}_{2}(z) S_{2}(z) + K^{+}_{3}(z)\log(z) \right.\nonumber\\
&& \left. + \log^{2}(z)K^{+}_{4}(z) + \log(z)\log(1-z)K^{+}_{5}(z)\right] A^{+}_{n}(y) +  \nonumber\\ 
&&K^{+}_{6}\left\{\int^{1}_{x}\frac{dy}{y} \frac{yA^{+}_{n}(y) - xA^{+}_{n}(x)}{y-x} + A^{+}_{n}(x)\log(1-x) \right\} + K^{+}_{7}\int^{1}_{x}\frac{dy}{y}A^{+}_{n}(y)\,, 
\eeqn

where $z={x}/{y}$. For the $(-)$ case we get a similar expression.
  
\begin{figure}[tbh]
{\centering \resizebox*{12cm}{!}{\rotatebox{-90}{\includegraphics{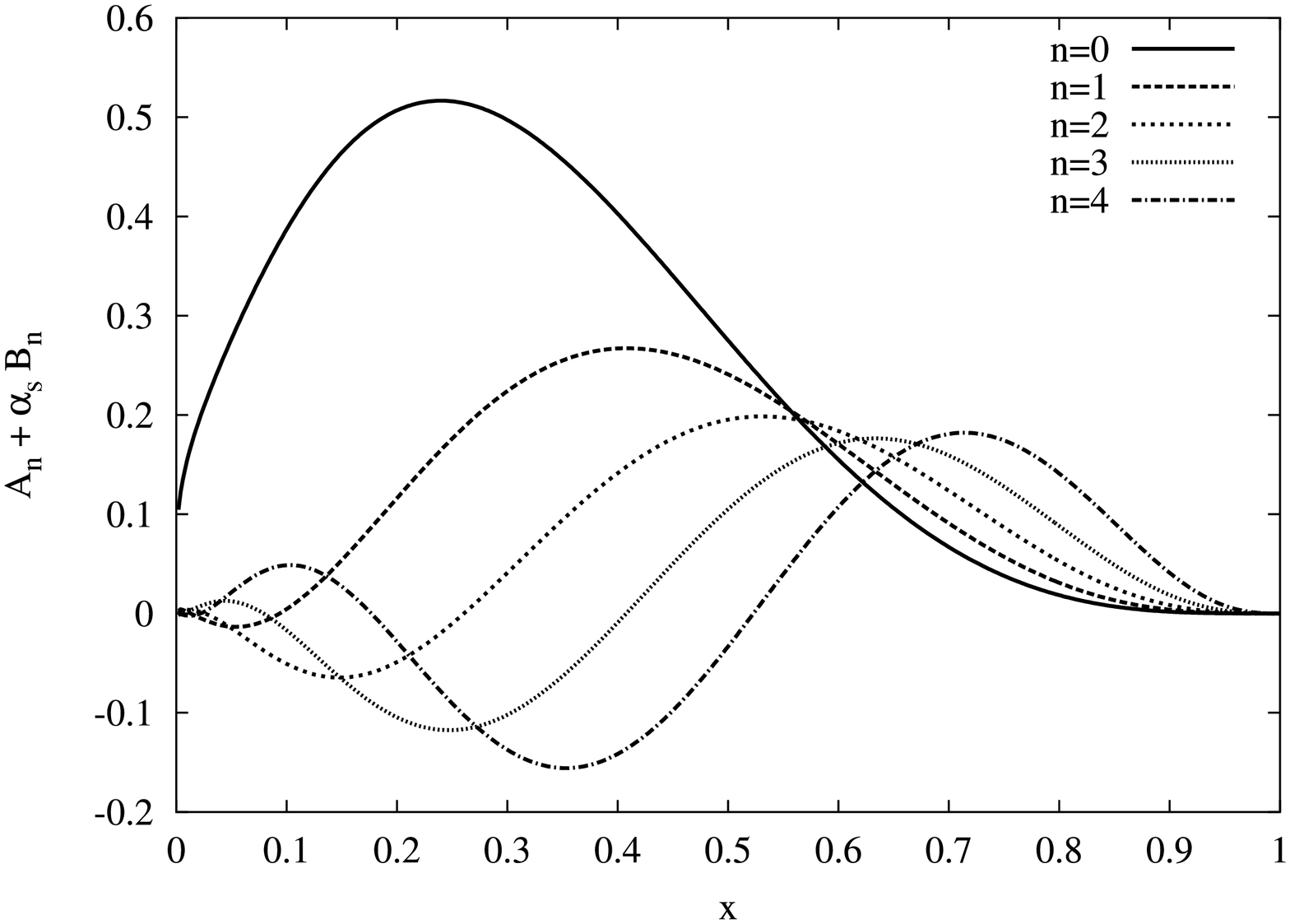}}} \par}
\caption{Coefficients \protect\( A_{n}(x)+\alpha _{s}(Q^{2})B_{n}\protect \),
with \protect\( n=0,\ldots ,4\protect \) for a final scale \protect\( Q=100\protect \)
GeV for the quark up.}
\label{an}
\end{figure}

\begin{figure}[tbh]
{\centering \resizebox*{12cm}{!}{\rotatebox{-90}{\includegraphics{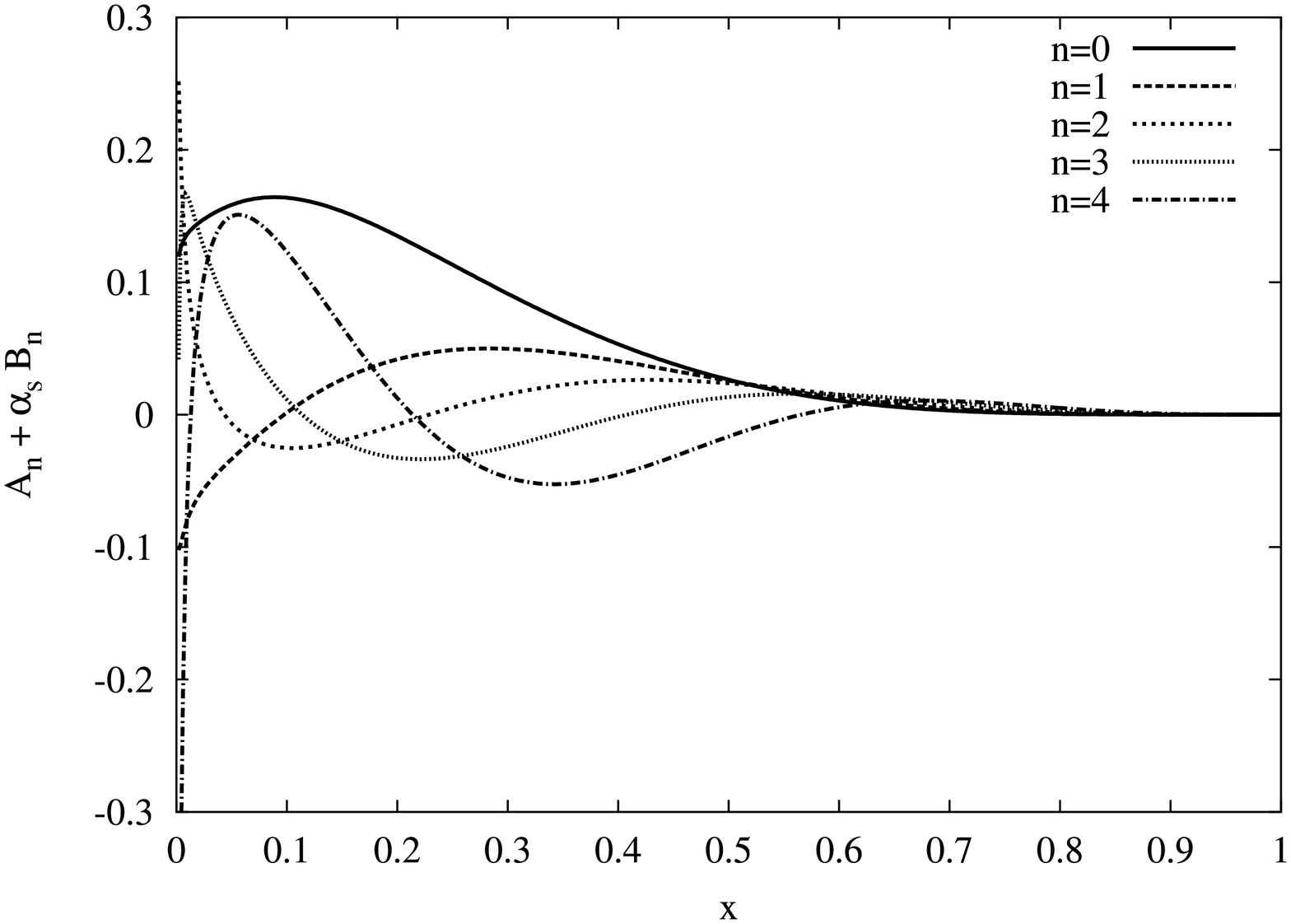}}} \par}

\caption{Coefficients \protect\( A_{n}(x)+\alpha _{s}(Q^{2})B_{n}\protect \),
with \protect\( n=0,\ldots ,4\protect \) for a final scale \protect\( Q=100\protect \)
GeV for the quark down.}
\label{anprime}
\end{figure}

Now we are ready to write down the expression for the $B^{\pm}_{n+1}(x)$ coefficient to NLO, 
similarly to eq. (\ref{recur1}). So we get (for the $(+)$ case) 

\ba
&&B^{+}_{n+1}(x) = - B^{+}_{n}(x) + \frac{\beta_{1}}{2\beta^{2}_{0}} \left\{2C_{F}\left[\int^{1}_{x}\frac{dy}{y}\frac{y A^{+}_{n}(y) - x A^{+}_{n}(x)}{y-x} + A^{+}_{n}(x) \log(1-x)\right]\right.+\nonumber\\
&&\left.-2C_{F}\left(\int_{x}^{1}\frac{dy}{y} A^{+}_{n}(y)\right) + C_{F}\frac{3}{2} A^{+}_{n}(x)\right\}-\frac{1}{4\pi\beta_{0}}K^{+}_{1} A^{+}_{n}(x)+ \int^{1}_{x}\frac{dy}{y}\left[ K^{+}_{2}(z) S_{2}(z) + \right.\nonumber\\
&&+ \left.K^{+}_{3}(z)\log(z)+\log^{2}(z)K^{+}_{4}(z) + \log(z)\log(1-z)K^{+}_{5}(z)\right]\left(-\frac{1}{4\pi\beta_{0}}\right)A^{+}_{n}(y)+\nonumber\\
&&K^{+}_{6}\left(-\frac{1}{4\pi\beta_{0}}\right)\left\{\left[\int^{1}_{x}\frac{dy}{y} \frac{yA^{+}_{n}(y) - xA^{+}_{n}(x)}{y-x} + A^{+}_{n}(x)\log(1-x) \right] + K^{+}_{7}\int^{1}_{x}\frac{dy}{y}A^{+}_{n}(y)\right\}-\nonumber\\
&&C_{F}\left(-\frac{4}{\beta_{0}}\right)\left[\int^{1}_{x}\frac{dy}{y}\frac{y B^{\pm}_{n}(y) - x B^{\pm}_{n}(x)}{y-x} + B^{\pm}_{n}(x) \log(1-x)\right]+\nonumber\\
&&C_{F}\left(\frac{4}{\beta_{0}}\right) \left(\int_{x}^{1}\frac{dy}{y} B^{\pm}_{n}(y)\right) + C_{F}\left(-\frac{2}{\beta_{0}}\right)\frac{3}{2} B^{\pm}_{n}(x)\,.\nonumber\\
\ea
As we have already mentioned, the implementation of these recursion relations require particular numerical care, since, as $n$ increases, numerical instabilities tend to add up 
unless high accuracy is used in the computation of the integrals. In particular we use finite 
element expansions to extract analitically the logarithms in the convolution 
(see the discussion in Appendix A). 
NLO plots of the coefficients $A_n(x) + \alpha(Q^2) B_n(x)$ are shown in figs. 
(\ref{an},\ref{anprime}) for a specific set of initial conditions (GRSV, as discussed below). 
As the index $n$ increases, the number of nodes also increases. A stable implementation 
can be reached for several thousands of grid-points and up to $n\approx 10$. 
Notice that the asymptotic expansion is suppressed by $n!$ and that additional 
contributions ($n>10$) are insignificant even at large ($> 200$ GeV) final evolution scales $Q$.

\section{Nonforward Extensions}
In this section we finally discuss the nonforward extension of the evolution 
algorithm. In the case of nonforward distributions a second scaling parameter $\zeta$ 
controls the asymmetry between the initial and the final nucleon momentum in the 
deeply virtual limit of nucleon Compton scattering. The solution of the evolution 
equations, in this case, are known in operatorial form. Single and double parton distributions are obtained sandwiching the operatorial solution with 4 possible types of initial/final states $<p|...|p>, <p|...|0>, <p'|...|p>$, corresponding, respectively, 
to the case of diagonal parton distributions, distribution amplitudes and, in the latter 
case, skewed and double parton distributions \cite{Radyushkin}. Here we will simply analize the non-singlet case and discuss the extension of the forward algorithm to this more general case. 
Therefore, given the off-forward distributions $H_q(x,\xi)$, in Ji's notation, 
we set up the expansion 
\beq 
H_q(x,\xi)=\sum_{k=0}^{\infty} \frac{A_k(x,\xi)}{k!}\log^k
\left(\frac{\alpha(Q^2)}{\alpha(Q_0^2)}\right) +
\alpha(Q^2)\sum_{k=0}^\infty \frac{B_k(x,\xi)}{k!}\log^k
\left(\frac{\alpha(Q^2)}{\alpha(Q_0^2)}\right),
\label{expansionx}
 \eeq
which is the natural extension of the forward algorithm discussed 
in the previous sections. We recall that in the light-cone gauge $H(x,\xi)$ 
is defined as 
\beq
H_q(x,\xi,\Delta^2))= \frac{1}{2}\int \frac{dy^-}{2 \pi}e^{-i x \bar{P}^+y^-}
\langle P'| \bar{\psi}_q(0,\frac{y^-}{2},{\bf 0_\perp})
\frac{1}{2}\gamma^+ \psi_q (0,\frac{y^-}{2},{\bf 0_\perp})|P\rangle
\eeq
with $\Delta= P' - P$, $\bar{P}^+=1/2(P + \bar{P})$ \cite{Ji} (symmetric choice) and 
$\xi \bar{P}=1/2\,\,\Delta^+$.

This distribution describes for $x>\xi$ and $x < -\xi$ the DGLAP-type region for the quark and the antiquark distribution respectively, and the ERBL 
\cite{ERBL} (see also \cite{CL} for an overview) distribution amplitude 
for $-\xi <x < \xi$. In the following we will omit the $\Delta$ dependence from $H_q$.  

Again, once we insert the ansatz (\ref{expansionx}) into the evolution equations we obtain an infinite set of recursion relations which we can solve numerically. In LO, it is rather simple to relate the Gegenbauer moments of the skewed distributions and those of the generalized 
scaling coefficients $A_n$.
We recall that in the 
nonforward evolution, the multiplicatively renormalizable operators appearing 
in the light cone expansion are given in terms of Gegenbauer polynomials 
\cite{Radyushkin}. 
The Gegenbauer moments of the coefficients $A_n$ of our expansion (\ref{expansionx}) 
can be easily related to those of the off-forward distribution 

\beq
C_n(\xi, Q^2) =\zeta^n\int_{-1}^{1} C_n^{3/2}(z/\xi)H(z,\xi, Q^2) dz. 
\eeq  
The evolution of these moments is rather simple  
\beq
C_n(\zeta, Q^2)=C_n(\zeta, Q_0^2) \left(\frac{\alpha(Q^2)}{\alpha(Q_0^2)}\right)^{\gamma_n/\beta_0} 
\eeq
with
\beq
\gamma_n= C_F \left(\frac{1}{2} - \frac{1}{(n+1)(n+2)} + 2 \sum_{j=2}^{n+1}\frac{1}{j}\right)
\eeq
being the non-singlet anomalous dimensions. If we define the Gegenbauer moments of 
our expansion 
\beq
A_k^{(n)}(\xi, Q^2) =\xi^n\int_{-1}^{1} C_n^{3/2}(z/\xi)H(z\,\xi, Q^2) dz 
\eeq  
we can relate the moments of the two expansions as 
\beq
A_k^{(n)}(\xi)=C_n(\zeta,Q_0^2)\left( \frac{\gamma_n}{\beta_0}\right)^k. 
\eeq
Notice that expansions similar to (\ref{expansionx}) hold also for other choices 
of kinematical variables, such as those defining the non-forward distributions 
\cite{Radyushkin}, where the t-channel longitudinal momentum exchange $\Delta^+$ is related to the longitudinal momentum of the incoming nucleon as $\Delta=\zeta P$. We recall 
that $H_q(x.\xi)$ as defined in \cite{Ji} can be mapped 
into two independent distributions $\hat{\mathcal{F}}_q(X,\zeta)$ and 
$\hat{\mathcal{F}}_{\bar{q}}(X,\zeta)$ through the mappings \cite{Golec}
\beqa
X_1 &=& \frac{(x_1+ \xi)}{(1 +\xi)} \nonumber \\
X_2 &=& \frac{\xi - x_2}{(1 +\xi)} \nonumber \\
\xi &=&\zeta/(2 - \zeta) \nonumber \\
\mathcal{F}_q(X_1,\zeta) &=& \frac{1}{1 - \zeta/2}H_q(x_1,\xi) \nonumber \\
\mathcal{F}_{\bar{q}}(X_2,\zeta) &=& \frac{-1}{1 - \zeta/2}H_q(x_2,\xi),\nonumber \\
\eeqa
in which the interval  $-1 \leq x \leq 1$ is split 
into two coverings, partially overlapping (for $-\xi\leq x \leq \xi$, or ERBL region) 
in terms of the two variables $-\xi \leq x_1 \leq 1$ ($0\leq X_1 \leq 1$) and 
$-1 \leq x_2 \leq \xi$ ($0\leq X_2 \leq 1$). In this new parameterization, the 
momentum fraction carried by the emitted quark is $X$, 
as in the case of ordinary distributions, where it is parametrized by Bjorken $x$. 
For definitess, we focus here on the DGLAP-like $(X> \zeta)$ region of 
the non-singlet evolution. The non-singlet kernel is given in this case by 
$(x\equiv X)$
\beqa
P_\zeta(x,\zeta)=\frac{\alpha}{\pi}C_F\left( \frac{1}{y - x}\left[1 + \frac{x x'}{y y'}\right] -
\delta(x - y)\int_0^1 dz \frac{1 + z^2}{1 - z}\right),
\label{nfkernel}
\eeqa
 we introduce a LO ansatz 
\beq 
\mathcal{F}_q(x,\zeta)=\sum_{k=0}^{\infty} \frac{\mathcal{A}_k(x,\zeta)}{k!}\log^k
\left(\frac{\alpha(Q^2)}{\alpha(Q_0^2)}\right)
\eeq
and insert it into the evolution of this region to obtain the very simple recursion relations 
\beqa
\A_{n+1}(X,\zeta) &=& -\frac{2}{\beta_0} C_F 
\int_X^1 \frac{dy}{y} \frac{y \A_n(y,\zeta) -
x \A_n(X,\zeta)}{y-X}  -\frac{2}{\beta_0} C_F 
\int_X^1 \frac{dy (X-\zeta)}{y(y-\zeta)} \frac{\left(y \A_n(X,\zeta) -X \A_n(y,\zeta)\right)}{y-X} 
\nonumber \\
&& -\frac{2}{\beta_0}C_F\hat{\A}_n(X,\zeta)\left[\frac{3}{2}+\ln\frac{(1 - X)^2(1 - x/\zeta)}{1 - \zeta}\right].
\eeqa
The recursion relations can be easily reduced to a weighted sum of contributions in which $\zeta$ is a spectator parameter. Here we will not make a complete implementation, but we will illustrate 
in an appendix the general strategy to be followed. There we show a very accurate analytical 
method to evaluate the logarithms generated by the expansion without having 
to rely on brute-force computations.  
 
\section{Positivity of the non-singlet Evolution}
Positivity of the non-singlet evolution is a simple consequence of the master-form associated to the 
non-forward kernel (\ref{nfkernel}). As we have already emphasized above, 
positivity of the initial conditions are sufficient to guarantee a positivity of the 
solution at any scale $Q$. The master-form of the equation allows to reinterpret the parton dynamics as a random walk biased toward small-x values as $\tau=\log(Q^2)$ 
increases.

In the non-forward case the identification of a transition probability 
for the random walk \cite{CafaCor} associated with the evolution of the parton distribution is obtained 
via the non-forward transition probability 
\beqa
w_\zeta(x|y) &=&\frac{\alpha}{\pi}C_F \frac{1}{y- x}\left[1 + \frac{x}{y}\frac{(x-\zeta)}{y - \zeta}\right]
\theta(y>x)\nonumber \\
w'_\zeta(y|x)&=&\frac{\alpha}{\pi}C_F \frac{x^2 + y^2}{x^2(x - y)}\theta(y<x)
\eeqa
and the corresponding master equation is given by 
\beq
\frac{d \mathcal{F}_q}{d\tau}=\int_x^1 dy\, w_\zeta(x|y)\mathcal{F}_q(y,\zeta,\tau)-
\int_0^x dy\, w'_\zeta(y|x)\mathcal{F}_q(x,\zeta,\tau),
\eeq
that can be re-expressed in a form which is a simple generalization of the formula for the 
forward evolution \cite{CafaCor}

\beqa
\frac{d \mathcal{F}_q}{d\log Q^2} &=& \int_x^1 dy\, w_\zeta(x|y)\mathcal{F}_q(y,\zeta,\tau) - 
\int_0^x dy \,w'_\zeta(y|x) \mathcal{F}_q(x,\zeta,\tau)
\nonumber \\
&=& -\int_0^{\alpha(x)} dy w_\zeta(x+y|x)* \mathcal{F}_q(x,\zeta,\tau)+ 
\int_0^{-x} dy\, w'_\zeta(x+y|x)\mathcal{F}_q(x,\zeta,\tau),
\eeqa
where a Moyal-like product appears 
\beq
w_\zeta(x+y|x)*\mathcal{F}_q(x,\zeta,\tau)\equiv w_\zeta(x+y|x) e^{-y \left(\overleftarrow{\partial}_x + 
\overrightarrow{\partial}_x\right)} \mathcal{F}_q(x,\zeta,\tau)
\eeq
and $\alpha(x) =x-1$. 
A Kramers-Moyal expansion of the equation allows to generate a differential equation 
of infinite order with a parametric dependence on $\zeta$ 

\beqa
\frac{d \mathcal{F}_q}{d\log Q^2} &=&\int_{\alpha(x)}^{0}dy\,  
w_\zeta(x+y|x)\mathcal{F}_q(x,\zeta,\tau) + 
\int_{0}^{-x}dy\,  
w'_\zeta(x+y|x)\mathcal{F}_q(x,\zeta,\tau) \nonumber \\
&& - \sum_{n=1}^{\infty}\int_0^{\alpha(x)}dy \frac{(-y)^n}{n!}{\partial_x}^n
\left(w_\zeta(x+y|x)\mathcal{F}_q(x,\zeta,\tau)\right).
\label{expans}
\eeqa
We define
\beqa
\tilde{a}_0(x,\zeta) &=& \int_{\alpha(x)}^{0} dy w_\zeta(x+y|x)\mathcal{F}_q(x,\zeta,\tau)
+ \int_{0}^{-x}dy\,  
w'_\zeta(x+y|x)\mathcal{F}_q(x,\zeta,\tau)
 \nonumber \\
a_n(x,\zeta) &=&\int_0^{\alpha(x)} dy \,y^n w_\zeta (x+y|x) \mathcal{F}_q(x,\zeta,\tau) \nonumber \\
\tilde{a}_n(x,\zeta)&=&\int_0^{\alpha(x)}
dy y^n {\partial_x}^n \left(w_\zeta(x+y|x)\mathcal{F}_q(x,\zeta,\tau)\right) \,\,\,n=1,2,...
\eeqa
If we arrest the expansion at the first two terms $(n=1,2)$ we are able to derive an 
approximate equation describing the dynamics of partons for non-diagonal transitions. 
The procedure is a slight generalization of the method presented in \cite{CafaCor}, 
to which we refer for further details. For this purpose we use the identities

\beqa
\tilde{a}_1(x,\zeta) &=&\partial_x a_1(x,\zeta) - \alpha(x) \partial_x \alpha(x)
w_\zeta(x + \alpha(x)|x)\mathcal{F}_q(x,\zeta,\tau) \nonumber \\
\tilde{a}_2(x,\zeta) &=&\partial_x^2 a_2(x,\zeta) - 
2 \alpha(x) (\partial_x \alpha(x))^2 w_\zeta(x+ \alpha(x)|x)\mathcal{F}_q(x,\zeta,\tau)
\nonumber \\
&&  - 
\alpha(x)^2 \partial_x\alpha(x)
\partial_x\left( w_\zeta(x+ \alpha(x)|x)\mathcal{F}_q(x,\zeta,\tau)\right)\nonumber \\ 
&& - \alpha^2(x)\partial_x \alpha(x) \partial_x\left( w_\zeta(x + y|x)
\mathcal{F}_q(x,\zeta,\tau)\right)|_{y=\alpha(x)}.
\eeqa
which allow to compute the first few coefficients of the expansion. 
Using these relations, the Fokker-Planck approximation to this equation 
can be worked out explicitely. We omit details on the derivation which is unobvious 
since particular care is needed to regulate the (canceling) divergences and just quote 
the result. 

A lengthy computation gives 
\beqa
\frac{d \mathcal{F}_q}{d\tau}&=& \frac{\alpha}{\pi}C_F\left(\frac{x_{0,-3}}{(x-\z)^3}
+ \frac{x_{0,-1}}{(x-\z)} + x_{0,0}\right)\mathcal{F}_q(x,\z,\tau) 
\nonumber \\
&& + \frac{\alpha}{\pi}C_F\left(\frac{x_{1,-3}}{(x-\z)^3}
+ \frac{x_{1,-1}}{(x-\z)}\right)\partial_x\mathcal{F}_q(x,\z,\tau) 
+ \frac{\alpha}{\pi}C_F\frac{x_{0,-3}}{(x-\z)^3}\partial_x^2\mathcal{F}_q(x,\z,\tau) 
\nonumber \\	
\eeqa
where we have defined  
\beqa
x_{0,-3}&=&\frac{-\left( {\left( -1 + x \right) }^3\,
      \left( 17 x^3 - \z^2 \left( 3 + 4\z \right)  + 
        3 x \z\left( 3 + 5 \z \right)  - 3 x^2\left( 3 + 7 \z \right) 
        \right)  \right) }{12\,x^3}\nonumber \\
x_{0,-1} &=&
\frac{-29 x^4 - 3 + x^2\,\left( -1 + \z \right)  + 2 \z - 
    2 x \left( 1 + 3 \z \right)  + x^3 \left( 12 + 23 \z \right) }{3 x^3}
\nonumber \\
x_{0,0} &=& 4 + \frac{1}{2 x^2} - \frac{3}{x} + 2 \log \frac{(1 - x)}{x}
\nonumber \\
x_{1,-1}&=& 
\frac{-\left( \left( -1 + 6 x - 15 x^2 + 14 x^3 \right) \,
      \left( x - \z \right)  \right) }{3 x^2}\nonumber \\
x_{1,-3}&=&
\frac{1}{2} - \frac{5 x}{3} + 5 x^3 - \frac{23 x^4}{6} + \frac{7 \z}{3} - 
  \frac{3 \z}{4 x} + \frac{5 x \z}{2} \nonumber \\
&& - 15 x^2 \z + 
  \frac{131 x^3 \z}{12} - \frac{5 \z^2}{2} + \frac{\z^2}{4 x^2} - 
  \frac{\z^2}{x} + 13 x \z^2 - \frac{39 x^2 \z^2}{4} - 3 \z^3 + 
  \frac{\z^3}{3 x^2} + \frac{8 x \z^3}{3} \nonumber \\
x_{2,-3}&=&\frac{-\left( {\left( -1 + x \right) }^2\,{\left( x - \z \right) }^2\,
      \left( 3 + 23 x^2 + 4 \z - 2 x\left( 7 + 8 \z \right)  \right) 
      \right) }{24 x}.
\eeqa

This equation and all the equations obtained by arresting the 
Kramers-Moyal expansion to higher order provide a complementary 
description of the non-forward dynamics in the DGLAP region, at least 
in the non-singlet case. Moving to higher order is straightforward 
although the results are slightly lengthier. A full-fledged 
study of these equations is under way and we expect that the DGLAP dynamics 
is reobtained - directly from these equations - as the order of the approximation increases.

\section{Model Comparisons, Saturation and the Tensor Charge}

In this last section we discuss some implementations of our methods to the 
standard (forward) evolution by doing a NLO model comparisons both in the 
analysis of Soffer's inequality and for the evolution of the tensor charge. 
We have selected two models, motivated quite independently 
and we have compared the predicted evolution of the Soffer bound at an 
accessable final evolution scale around $100$ GeV for the light quarks and 
around $200$ GeV for the heavier generations. At this point we recall that 
in order to generate suitable initial 
conditions for the analysis of Soffer's inequality, one needs an ansatz 
in order to quantify the difference between its left-hand side and right-hand side 
at its initial value.

The well known strategy to build reasonable initial conditions for the transverse 
spin distribution consists in generating polarized distributions (starting 
from the unpolarized ones) and then saturate the inequality at some lowest scale, 
which is the approach we have followed for all the models that we have implemented. 

Following Ref. \cite{GRSV} (GRSV model), 
we have used as input distributions - in the unpolarized case - the formulas
in Ref. \cite{GRV}, calculated to NLO in the \( \overline{\textrm{MS}} \)
scheme at a scale \( Q_{0}^{2}=0.40\, \textrm{GeV}^{2} \)
\begin{eqnarray}
x(u-\overline{u})(x,Q_{0}^{2}) & = & 0.632x^{0.43}(1-x)^{3.09}(1+18.2x)\nonumber \\
x(d-\overline{d})(x,Q_{0}^{2}) & = & 0.624(1-x)^{1.0}x(u-\overline{u})(x,Q_{0}^{2})\nonumber \\
x(\overline{d}-\overline{u})(x,Q_{0}^{2}) & = & 0.20x^{0.43}(1-x)^{12.4}(1-13.3\sqrt{x}+60.0x)\nonumber \\
x(\overline{u}+\overline{d})(x,Q_{0}^{2}) & = & 1.24x^{0.20}(1-x)^{8.5}(1-2.3\sqrt{x}+5.7x)\nonumber \\
xg(x,Q_{0}^{2}) & = & 20.80x^{1.6}(1-x)^{4.1}
\end{eqnarray}
and \( xq_{i}(x,Q_{0}^{2})=x\overline{q_{i}}(x,Q_{0}^{2})=0 \) for
\( q_{i}=s,c,b,t \).

We have then related the unpolarized input distribution to the longitudinally
polarized ones by as in Ref. \cite{GRSV} 
\begin{eqnarray}
x\Delta u(x,Q_{0}^{2}) & = & 1.019x^{0.52}(1-x)^{0.12}xu(x,Q_{0}^{2})\nonumber \\
x\Delta d(x,Q_{0}^{2}) & = & -0.669x^{0.43}xd(x,Q_{0}^{2})\nonumber \\
x\Delta \overline{u}(x,Q_{0}^{2}) & = & -0.272x^{0.38}x\overline{u}(x,Q_{0}^{2})\nonumber \\
x\Delta \overline{d}(x,Q_{0}^{2}) & = & x\Delta \overline{u}(x,Q_{0}^{2})\nonumber \\
x\Delta g(x,Q_{0}^{2}) & = & 1.419x^{1.43}(1-x)^{0.15}xg(x,Q_{0}^{2})
\end{eqnarray}
and \( x\Delta q_{i}(x,Q_{0}^{2})=x\Delta \overline{q_{i}}(x,Q_{0}^{2})=0 \)
for \( q_{i}=s,c,b,t \). 

Following \cite{MSSV}, we assume the saturation of
Soffer inequality:\begin{equation}
\label{eq:saturation}
x\Delta _{T}q_{i}(x,Q_{0}^{2})=\frac{xq_{i}(x,Q_{0}^{2})+x\Delta q_{i}(x,Q_{0}^{2})}{2}
\end{equation}
and study the impact of the different evolutions 
on both sides of Soffer's inequality at various final evolution scales $Q$. 

In the implementation of the second model (GGR model)
we have used as input distributions in the unpolarized case the 
CTEQ4 parametrization \cite{CTEQ}, calculated to NLO in the
\( \overline{\textrm{MS}} \) scheme at a scale \( Q_{0}=1.0\, \textrm{GeV} \)

\begin{eqnarray}
x(u-\overline{u})(x,Q_{0}^{2}) & = & 1.344x^{0.501}(1-x)^{3.689}(1+6.402x^{0.873})\nonumber \\
x(d-\overline{d})(x,Q_{0}^{2}) & = & 0.64x^{0.501}(1-x)^{4.247}(1+2.69x^{0.333})\nonumber \\
xs(x,Q_{0}^{2})=x\overline{s}(x,Q_{0}^{2}) & = & 0.064x^{-0.143}(1-x)^{8.041}(1+6.112x)\nonumber \\
x(\overline{d}-\overline{u})(x,Q_{0}^{2}) & = & 0.071x^{0.501}(1-x)^{8.041}(1+30.0x)\nonumber \\
x(\overline{u}+\overline{d})(x,Q_{0}^{2}) & = & 0.255x^{-0.143}(1-x)^{8.041}(1+6.112x)\nonumber \\
xg(x,Q_{0}^{2}) & = & 1.123x^{-0.206}(1-x)^{4.673}(1+4.269x^{1.508})
\end{eqnarray}
and \( xq_{i}(x,Q_{0}^{2})=x\overline{q_{i}}(x,Q_{0}^{2})=0 \) for
\( q_{i}=c,b,t \)
and we have related the unpolarized input distribution to the longitudinally
polarized ones by the relations \cite{GGR}

\begin{eqnarray}
x\Delta \overline{u}(x,Q_{0}^{2}) & = & x\eta _{u}(x)xu(x,Q_{0}^{2})\nonumber \\
x\Delta u(x,Q_{0}^{2}) & = & \cos \theta _{D}(x,Q_{0}^{2})\left[ x(u-\overline{u})-\frac{2}{3}x(d-\overline{d})\right] (x,Q_{0}^{2})+x\Delta \overline{u}(x,Q_{0}^{2})\nonumber \\
x\Delta \overline{d}(x,Q_{0}^{2}) & = & x\eta _{d}(x)xd(x,Q_{0}^{2})\nonumber \\
x\Delta d(x,Q_{0}^{2}) & = & \cos \theta _{D}(x,Q_{0}^{2})\left[ -\frac{1}{3}x(d-\overline{d})(x,Q_{0}^{2})\right] +x\Delta \overline{d}(x,Q_{0}^{2})\nonumber \\
x\Delta s(x,Q_{0}^{2})=x\Delta \overline{s}(x,Q_{0}^{2}) & = & x\eta _{s}(x)xs(x,Q_{0}^{2})
\end{eqnarray}
and \( x\Delta q_{i}(x,Q_{0}^{2})=x\Delta \overline{q_{i}}(x,Q_{0}^{2})=0 \)
for \( q_{i}=c,b,t \).

\begin{figure}[tbh]
{\centering \resizebox*{12cm}{!}{\rotatebox{-90}{\includegraphics{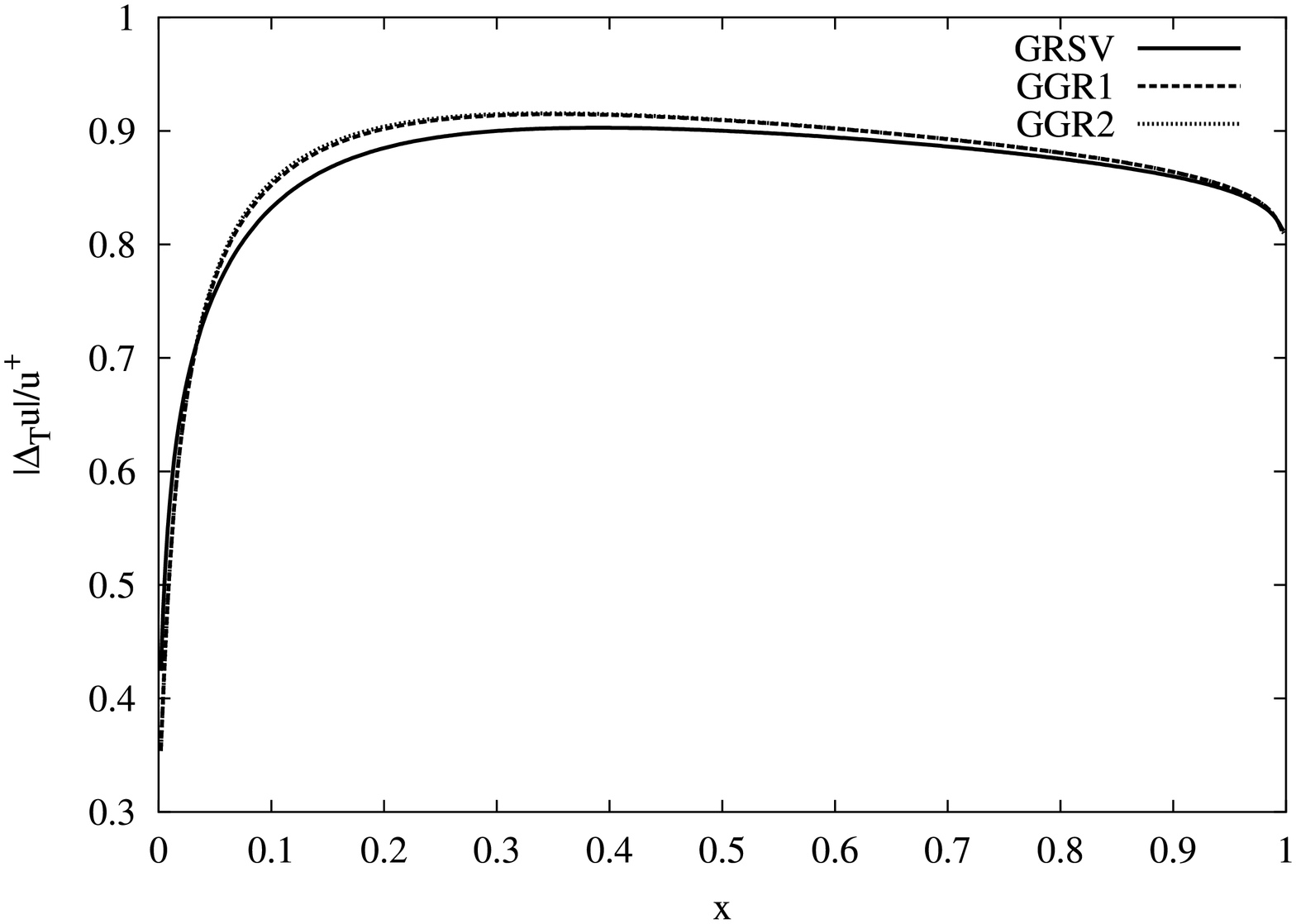}}} \par}
\caption{ Test of Soffer's inequality for quark up at \protect\( Q=100\protect \)
GeV for different models.}
\label{upsof}
\end{figure}

\begin{figure}[tbh]
{\centering \resizebox*{12cm}{!}{\rotatebox{-90}{\includegraphics{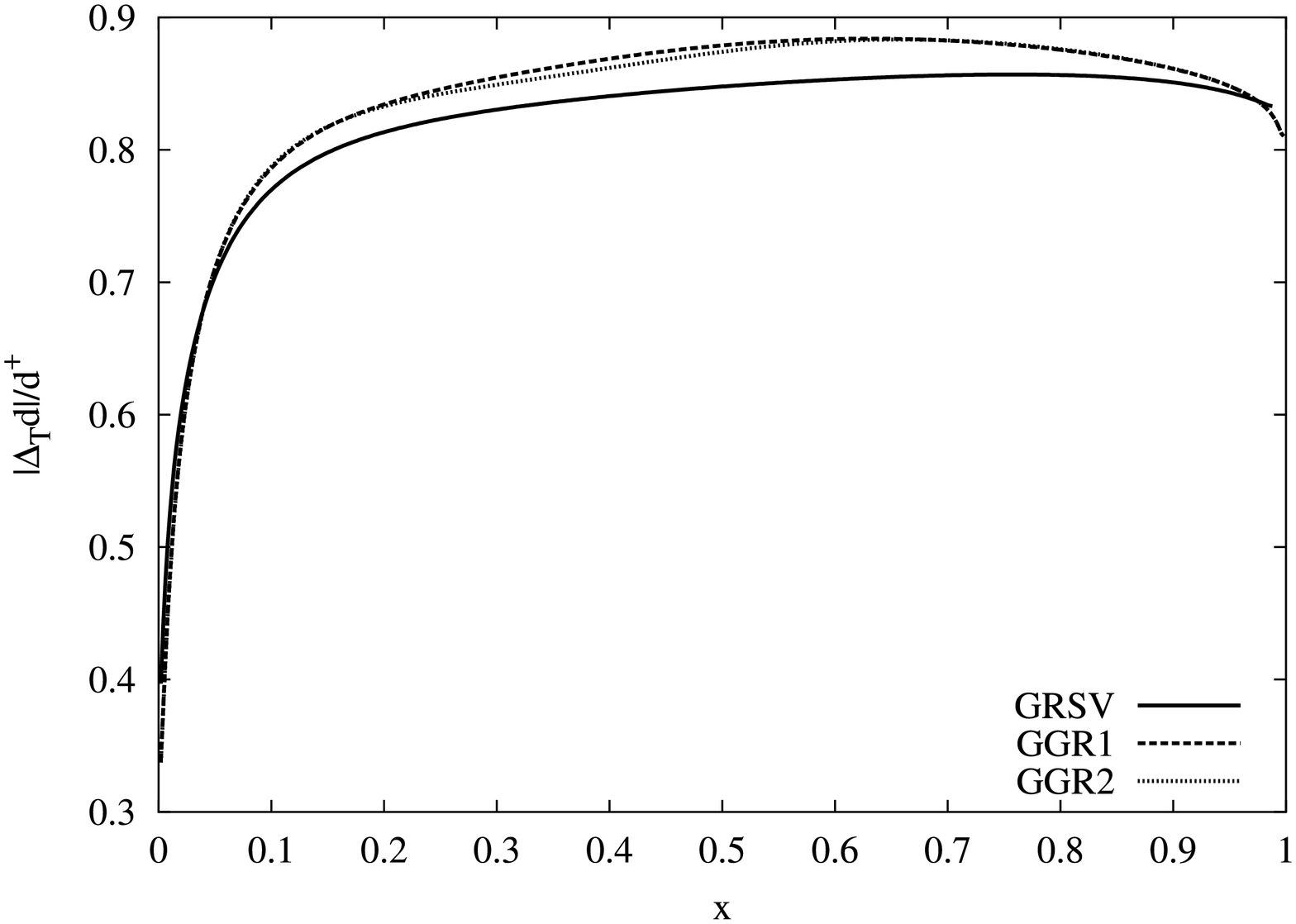}}} \par}
\caption{ Test of Soffer's inequality for quark down at \protect\( Q=100\protect \)
GeV for different models}
\label{downsof}
\end{figure}

\begin{figure}[tbh]
{\centering \resizebox*{12cm}{!}{\rotatebox{-90}{\includegraphics{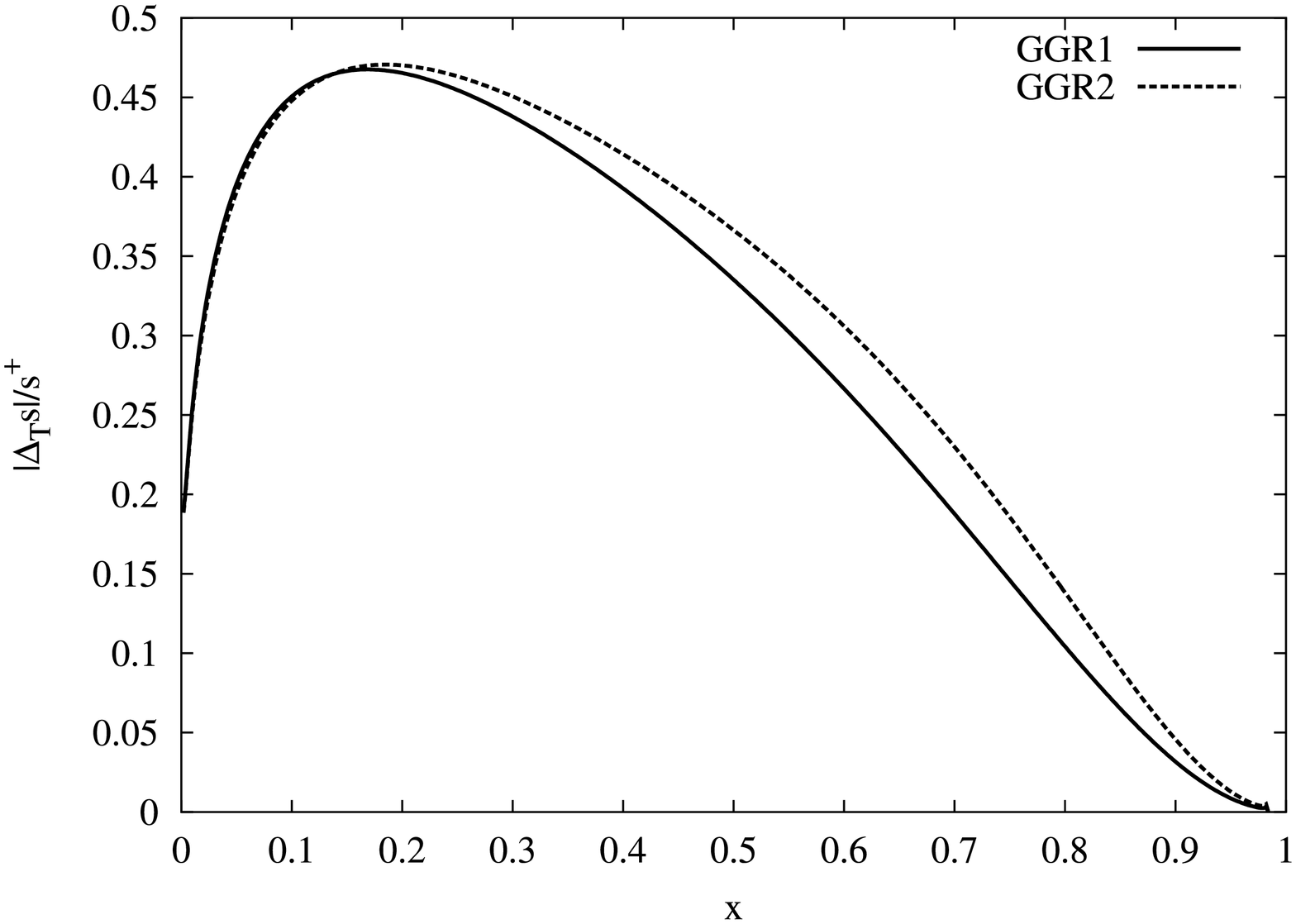}}} \par}
\caption{ Test od Soffer's inequality for quark strange at \protect\( Q=100\protect \)
GeV for different models}
\label{strangesof}
\end{figure}

\begin{figure}[tbh]
\resizebox*{9cm}{!}{\rotatebox{-90}{\includegraphics{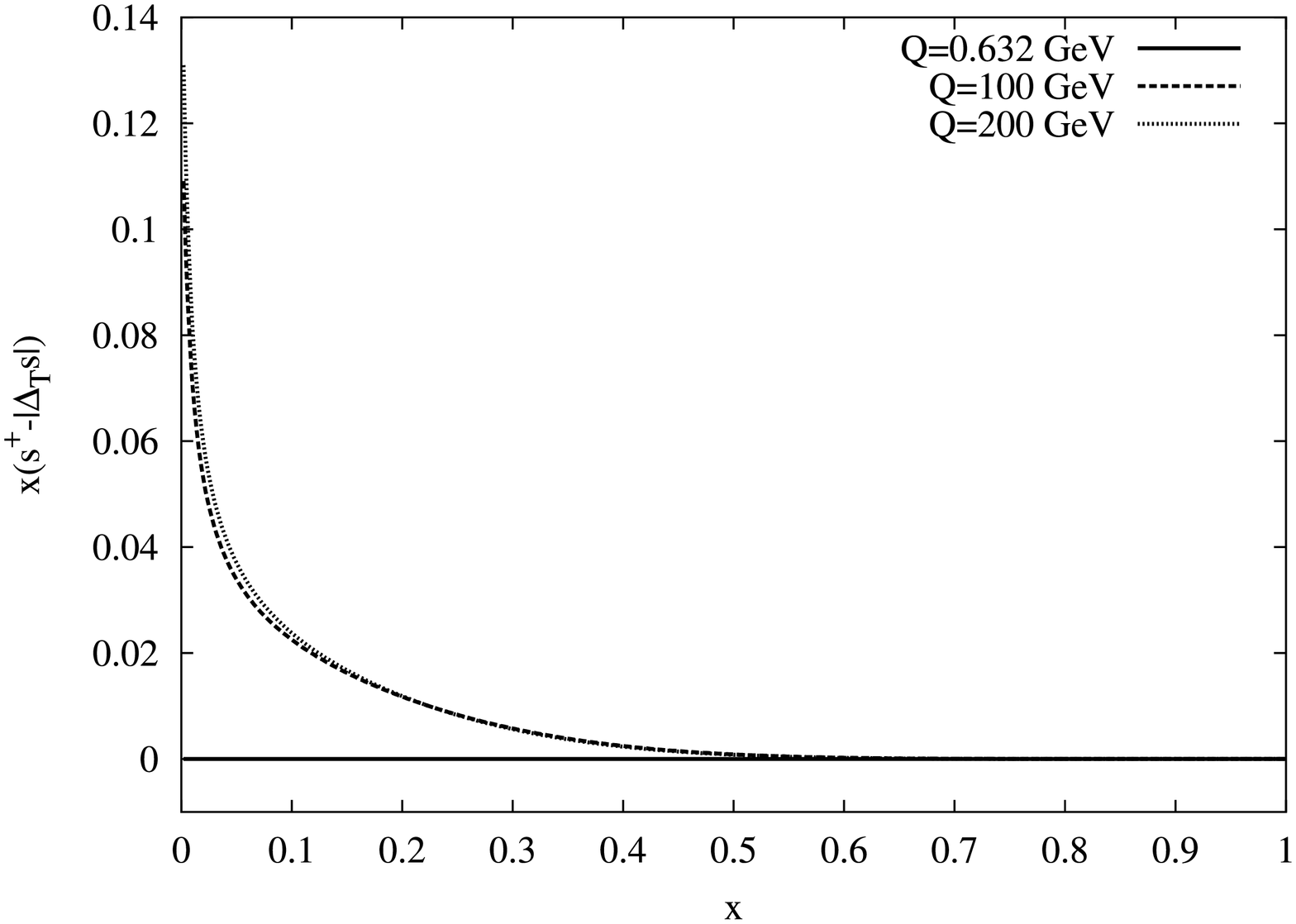}}} 
\resizebox*{9cm}{!}{\rotatebox{-90}{\includegraphics{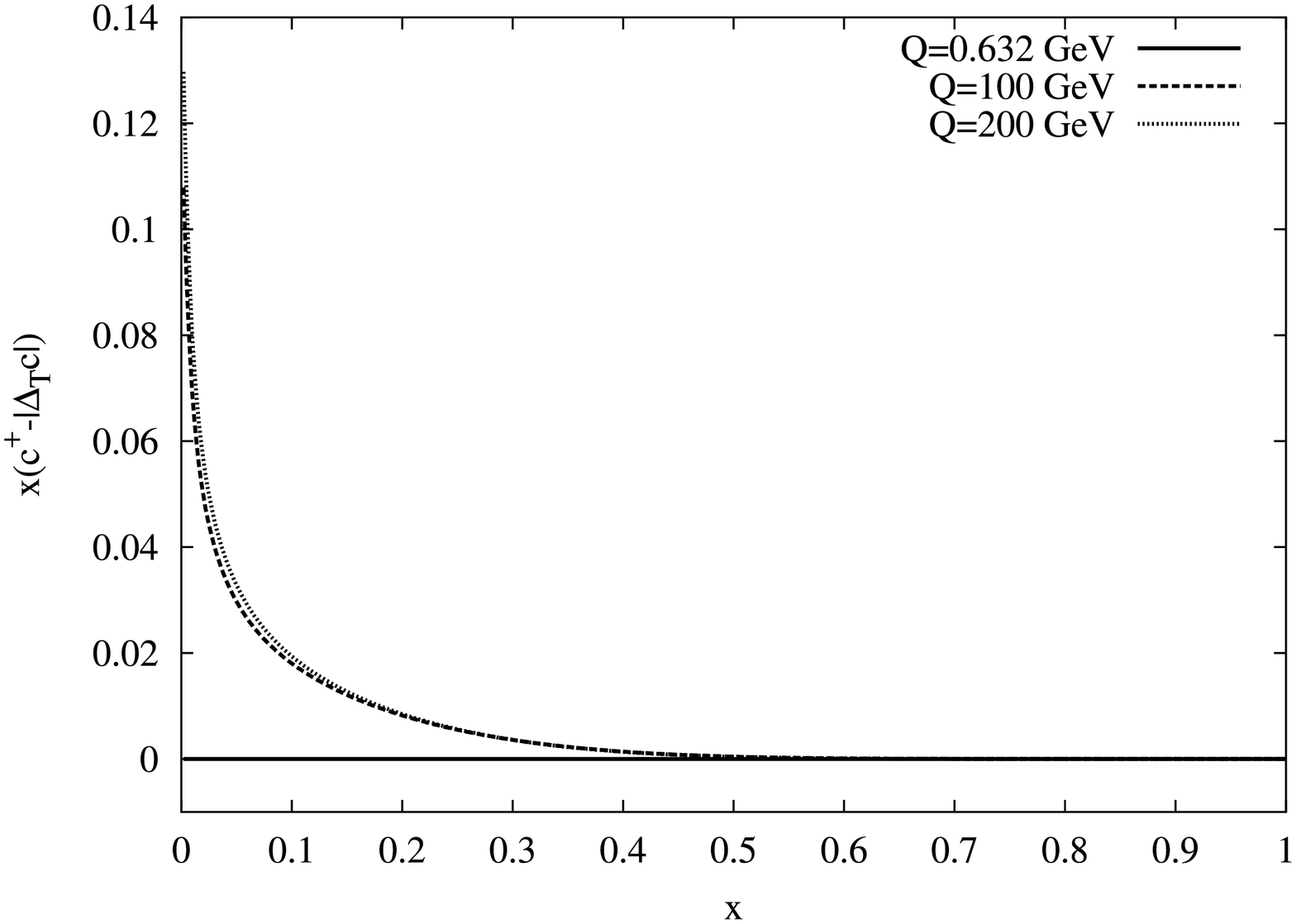}}} 
\caption{Soffer's inequality for strange and charm in the GRSV model.}
\label{scsof}
\end{figure}

\begin{figure}[tbh]
\resizebox*{9cm}{!}{\rotatebox{-90}{\includegraphics{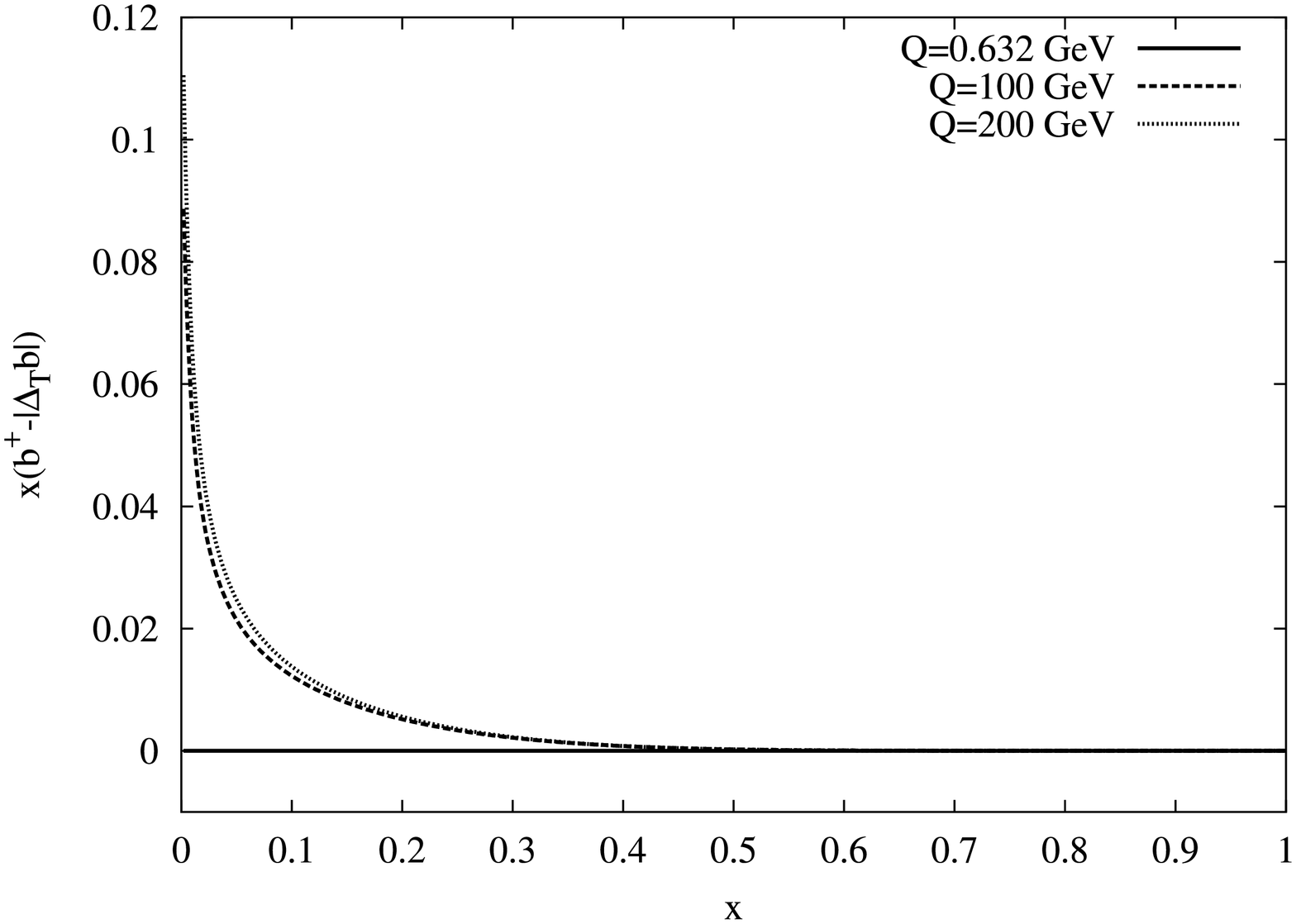}}} 
\resizebox*{9cm}{!}{\rotatebox{-90}{\includegraphics{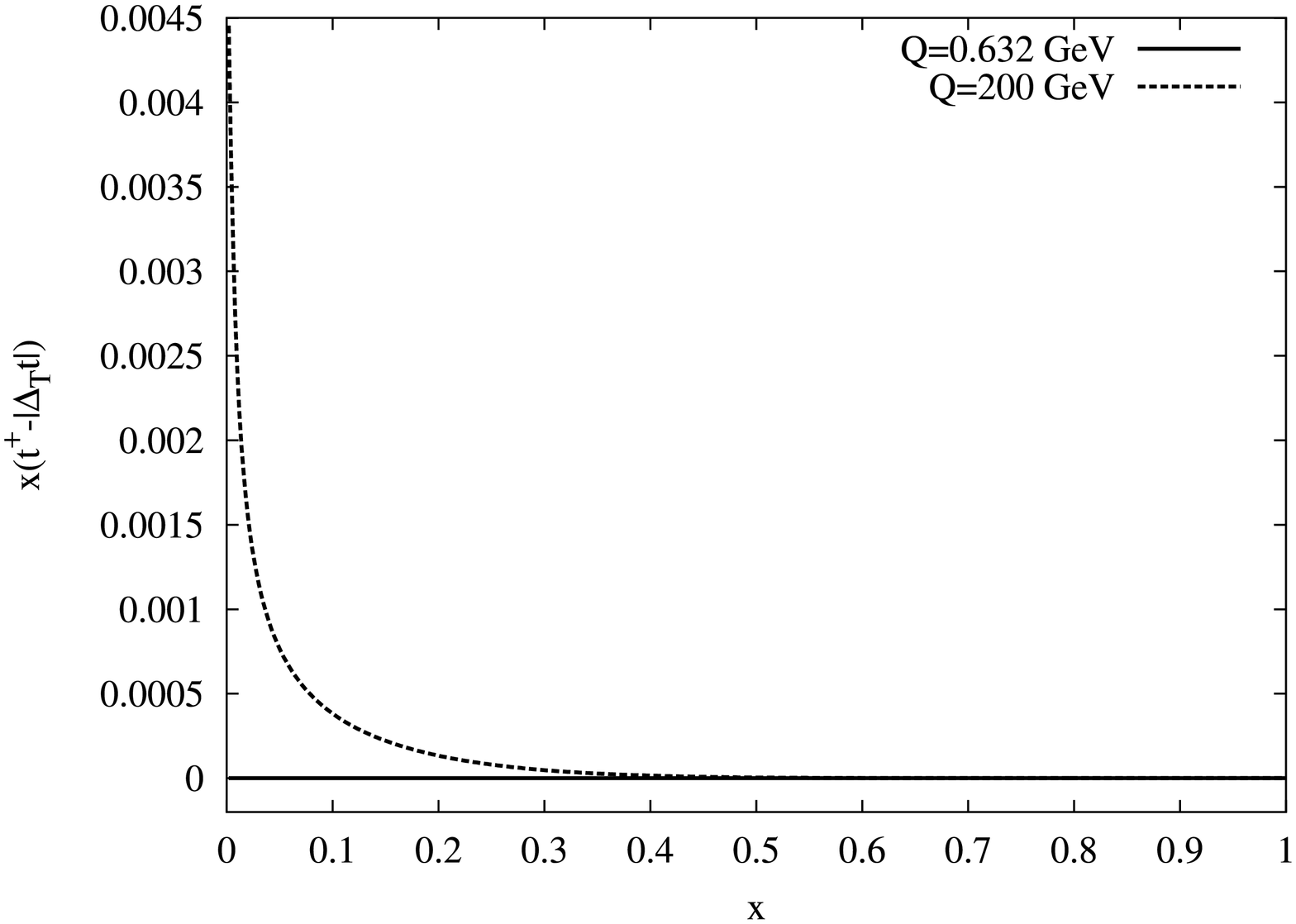}}} 
\caption{Soffer's inequality for bottom and top
in the GRSV model.}
\label{btsof}
\end{figure}

A so-called ``spin dilution factor'' as defined in \cite{GGR}, which appears 
in the equations above is given by
\begin{equation}
\cos \theta _{D}(x,Q_{0}^{2})=\left[ 1+\frac{2\alpha _{s}(Q^{2})}{3}\frac{(1-x)^{2}}{\sqrt{x}}\right] ^{-1}.
\end{equation}
In this second (GGR) model, in regard to the initial conditions for the gluons, 
we have made use of two different options, characterized by a parameter 
\( \eta  \) dependent on the corresponding option. 
The first option, that
we will denote by GGR1, assumes that gluons are moderately polarized 
\begin{eqnarray}
x\Delta g(x,Q_{0}^{2}) & = & x\cdot xg(x,Q_{0}^{2})\nonumber \\
\eta _{u}(x)=\eta _{d}(x) & = & -2.49+2.8\sqrt{x}\nonumber \\
\eta _{s}(x) & = & -1.67+2.1\sqrt{x},
\end{eqnarray}
while the second option (GGR2) assumes that gluons are not polarized 
\begin{eqnarray}
x\Delta g(x,Q_{0}^{2}) & = & 0\nonumber \\
\eta _{u}(x)=\eta _{d}(x) & = & -3.03+3.0\sqrt{x}\nonumber \\
\eta _{s}(x) & = & -2.71+2.9\sqrt{x}.
\end{eqnarray}
We have plotted both ratios $\Delta_T/f^+$ and differences 
$(x f^+ - x\Delta_T f)$ for 
various flavours as a function of $x$. For the up quark, while the two models GGR1 and GGR2 
are practically overlapping, the difference between the GGR 
and the GRSV models in the the ratio $\Delta_T u/u^+$ 
is only slightly remarked in the intermediate x region $(0.1-0.5)$. In any case, it is just at the few percent level (Fig. (\ref{upsof})), while the inequality is satisfied 
with a ratio between the plus helicity distribution and transverse around 10 percent from the saturation value, and above. There is a wider gap in the inequality at small x, region characterized by larger transverse distribution, with values up to 40 percent from saturation. A similar trend is noticed for the x-behaviour of the inequality in the case 
of the down quark (Fig. \ref{downsof}). In this latter case the GGR 
and the GRSV model show a more remarked difference, especially for 
intermediate x-values. An interesting features appears in the corresponding 
plot for the strange quark (Fig.(\ref{strangesof})), showing a much 
wider gap in the inequality (50 percent and higher) compared 
to the other quarks. Here we have plotted results for the two GGR models (GGR1 and GGR2). 
Differently from the case of the other quarks, in this case we observe 
a wider gap between lhs and rhs at larger x values, increasing as $x\rightarrow 1$. 
In figs. (\ref{scsof})and (\ref{btsof}) we plot the differences $(x f^+ - x\Delta_T f)$ 
for strange and charm and for bottom and top quarks respectively, which 
show a much more reduced evolution from the saturation value up to the final corresponding 
evolving scales (100 and 200 GeV). 
As a final application we finally discuss the behaviour of the tensor charge 
of the up quark for the two models as a function of the final 
evolution scale $Q$. We recall that like the isoscalar and the isovector 
axial vector charges defined from the forward matrix element 
of the nucleon, the nucleon tensor charge is defined from the matrix 
element of the tensor current 
\beq
\langle P S_T|\bar{\psi}\sigma^{\mu\nu}\gamma_5 \lambda^a \psi|P,S_T\rangle 
=2 \delta q^a(Q_0^2)\left( P^\mu S_T^\nu - P^\nu S_T^\mu\right)
\eeq
where $\delta^a q(Q_0^2)$ denotes the flavour (a) contribution to the nucleon tensor charge at a scale $Q_0$ and $S_T$ is the transverse spin. 

\begin{figure}[tbh]
{\centering \resizebox*{12cm}{!}{\rotatebox{-90}{\includegraphics{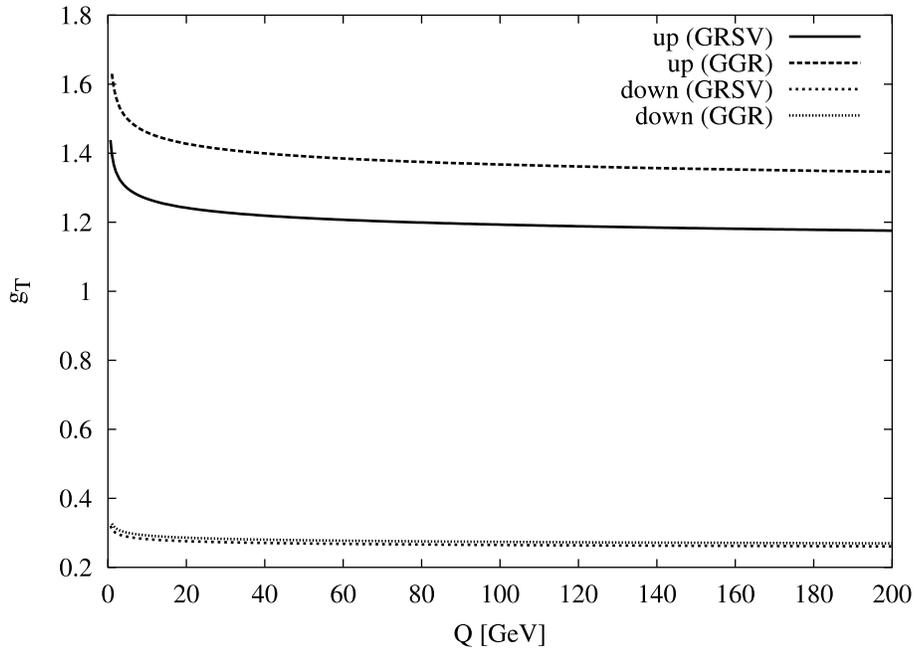}}} \par}
\caption{Tensor charge \protect\( g_{T}\protect \) as a function of \protect\( Q\protect \)
for up and down quark for the GRSV and GGR models.}
\label{figure}
\end{figure}

In fig. (\ref{figure}) we plot the evolution of the tensor charge for the models 
we have taken in exam. At the lowest evolution scales the charge is, in these models, 
above 1 and decreases slightly as the factorization scale $Q$ increases. 
We have performed an evolution up to 200 GeV as an illustration of this behaviour. 
There are substantial differences between these models, as one can easily observe, which are around 20 percent. From the analysis of these differences at various factorization 
scales we can connect low energy dynamics to observables at higher energy, thereby 
distinguishing between the various models. Inclusion of the correct evolution, 
up to subleading order is, in general, essential.

\section{Conclusions}
We have illustrated the use of x-space based algorithms for the solution 
of evolution equations in the leading and in the next-to-leading 
approximation and we have 
provided some applications of the method both in the analysis 
of Soffer's inequality and in the investigation 
of other relations, such as the evolution of 
the proton tensor charge, for various models. 
The evolution has been implemented using a suitable base, 
relevant for an analysis of positivity in LO, using kinetic arguments. 
The same kinetic argument 
has been used to prove the positivity of the evolution of 
$h_1$ and of the tensor charge up to NLO. 
In our implementations we have completely relied on recursion relations without 
any reference to Mellin moments. We have provided several illustrations of the 
recursive algorithm and extended it to the non-forward evolution up to NLO. 
Building on previous work 
for the forward evolution, we have presented a master-form of the non-singlet 
evolution of the skewed distributions, a simple proof of positivity
and a related Kramers Moyal expansion, 
valid in the DGLAP region of the skewed evolution for any value 
of the asymmetry parameter $\zeta$. We hope to return with a complete study of the 
nonforward evolution and related issues not discussed here in the near future.

\vspace{1cm}
\centerline{\bf Acknowledgements}
C.C. thanks Prof. Ph. Jetzer and the Theory Group at the University of Zurich 
for hospitality while completing this work. 

\section{Appendix A. Weighted Sums}
In this appendix we briefly illustrate the reduction of 
recursion relations to analytic expressions based on finite element decompositions of the 
corresponding integrals. The method allows to write in analytic forms the 
most dangerous integrals thereby eliminating possible sources of instabilities in the 
implementation of the recursion relations. The method uses a linear 
interpolation formula for the coefficients $A_n$, $B_n$ which, in principle can also 
be extended to higher (quadratic) order. However, enugh accuracy can be achieved by 
increasing the grid points in the discretization. Notice that using this method 
we can reach any accuracy since we have closed formulas for the integrals. 
In practice these and similar equations are introduced analytically as functions in the 
numerical integration procedures.  

Below, we will use a simplified notation ($X\equiv x$ for simplicity). 

We define $\bar{P}(x,\zeta)\equiv x P(x,\zeta)$ and $\bar{A}(x,\zeta)\equiv x A(x)$ and the 
convolution products 

\beq
J(x)\equiv\int_x^1 \frac{dy}{y}\left(\frac{x}{y}\right)
P\left(\frac{x}{y},\zeta\right)\bar{A}(y). \
\eeq
The integration interval in $y$ at any fixed x-value is
partitioned in an array of
increasing points ordered from left to right
$\left(x_0,x_1,x_2,...,x_n,x_{n+1}\right)$
with $x_0\equiv x$ and $x_{n+1}\equiv 1$ being the upper edge of the
integration
region. One constructs a rescaled array
$\left(x,x/x_n,...,x/x_2,x/x_1, 1 \right)$. We define
$s_i\equiv x/x_i$, and $s_{n+1}=x < s_n < s_{n-1}<... s_1 < s_0=1$.
We get
\beq
J(x,\zeta)=\sum_{i=0}^N\int_{x_i}^{x_{i+1}}\frac{dy}{y}
\left(\frac{x}{y}\right) P\left(\frac{x}{y},\zeta\right)\bar{A}(y,\zeta)
\eeq
At this point we introduce the linear interpolation
\beq
\bar{A}(y,\zeta)=\left( 1- \frac{y - x_i}{x_{i+1}- x_i}\right)\bar{A}(x_i,\zeta) +
\frac{y - x_i}{x_{i+1}-x_i}\bar{A,\zeta}(x_{i+1})
\label{inter}
\eeq
and perform the integration on each subinterval with a change of variable
$y->x/y$ and replace the integral $J(x,\zeta)$ with
its discrete approximation $J_N(x)$
to get
\beqa
J_N(x,\zeta) &=& \bar{A}(x_0)\frac{1}{1- s_1}\int_{s_1}^1 \frac{dy}{y}P(y,\zeta)(y - s_1)
\nonumber \\
&+& \sum_{i=1}^{N}\bar{A}(x_i,\zeta) \frac{s_i}{s_i - s_{i+1}}
\int_{s_{i+1}}^{s_i} \frac{dy}{y}P(y)(y - s_{i+1})\nonumber \\
& -& \sum_{i=1}^{N}\bar{A}(x_i,\zeta) \frac{s_i}{s_{i-1} - s_{i}}
\int_{s_{i}}^{s_{i-1}} \frac{dy}{y}P(y,\zeta)(y - s_{i-1}) \nonumber \\
\eeqa
with the condition $\bar{A}(x_{N+1},\zeta)=0$.
Introducing the coefficients  $W(x,x,\zeta)$ and $W(x_i,x,\zeta)$, the integral
is cast in the form
\beq
J_N(x,\zeta)=W(x,x,\zeta) \bar{A}(x,\zeta) + \sum_{i=1}^{n} W(x_i,x,\zeta)\bar{A}(x_i,\zeta)
\eeq
where
\beqa
W(x,x,\zeta) &=& \frac{1}{1-s_1} \int_{s_1}^1 \frac{dy}{y}(y- s_1)P(y,\zeta), \nonumber \\
W(x_i,x,\zeta) &=& \frac{s_i}{s_i- s_{i+1}}
\int_{s_{i+1}}^{s_i} \frac{dy}{y}\left( y - s_{i+1}\right) P(y,\zeta) \nonumber \\
& -& \frac{s_i}{s_{i-1} - s_i}\int_{s_i}^{s_{i-1}}\frac{dy}{y}\left(
y - s_{i-1}\right) P(y,\zeta).\nonumber \\
\eeqa
For instance, after some manipulations we get 
\beq
\int_X^1 \frac{dy}{y} \frac{y \A_n(y,\zeta) -
x \A_n(X,\zeta)}{y-X}= {\bf In_0}(x)\A_n(x,\zeta) +\sum_{i+1}^{N} 
\left({\bf Jn_i}(x)- {\bf Jnt_i}(x)\right) \A_n(x_i) - \ln( 1- x) \A_n(x,\zeta)
\eeq
where
\beqa
{\bf I_0}(x) & = &
 \frac{1}{1- s_1} \log(s_1) + \log(1- s_1) \nonumber \\
\nonumber \\
{\bf J_i}(x) & = & \frac{1}{s_i - s_{i +1}}
\left[ \log\left(\frac{1 - s_{i+1}}{1 - s_i}\right)
+ s_{i+1} \log\left(\frac{1- s_i}{1 - s_{i+1}}
\frac{s_{i+1}}{s_i}\right)\right]
\nonumber \\
{\bf J'_i}(x) & = & \frac{1}{s_{i-1}- s_i}\left[ \log\left(\frac{1 - s_i}{1 -
s_{i-1}}\right)
+ s_{i-1}\log\left( \frac{s_i}{s_{i-1}}\right) + s_{i-1}\left(
\frac{1 - s_{i-1}}{1 - s_i}\right)\right],   \,\,\,\,\,\ i=2,3,..N \nonumber \\
{\bf J_1}(x) &=& \frac{1}{1- s_1}\log s_1. \nonumber \\
\eeqa
These functions, as shown here, and similar ones, are computed once and for all 
the kernels and allow to obtain very fast and extremely accurate implementations
for any $\zeta$.

\section{Appendix B}

\makeatother

\begin{tabular}{|c|c|c|}
\hline
\( n_{f} \)&
\( A \)&
\( B \)\\
\hline
\hline
3&
12.5302&
12.1739\\
\hline
4&
10.9569&
10.6924\\
\hline
5& 
9.3836&
9.2109\\
\hline
6&
7.8103&
7.7249\\
\hline
\end{tabular}

Table 1. Coefficients A and B for various flavour, to NLO 
for $\Delta_T P_{qq, \pm}$.

\section{APPENDIX C}
Here we define some notations in regard to the recursion relations used for the 
NLO evolution of the transverse spin distributions. 

For the $(+)$ case we have these expressions 

\ba
&&K_{1}^{+}(x)=\frac{1}{72}C_{F} (-2 n_{f} (3+4\pi^{2}) + N_{C}(51 + 44\pi^{2} - 216 \zeta(3))+ 9C_{F}(3-4\pi^{2}+48\zeta(3))\nonumber\\
&&K_{2}^{+}(x)= \frac{2 C_{F}(-2C_{F}+N_{C})x}{1+x}\\
&&K_{3}^{+}(x)= \frac{C_{F}(9C_{F}-11 N_{C}+2n_{f})x}{3(x-1)}\\
&&K_{4}^{+}(x)=\frac{C_{F}N_{C}x}{1-x}\\ 
&&K_{5}^{+}(x)=\frac{4C_{F}^{2}x}{1-x}\\
&&K_{6}^{+}(x)=-\frac{1}{9}C_{F}(10n_{f}+N_{C}(-67 + 3\pi^{2}))\\
&&K_{7}^{+}(x)=\frac{1}{9}C_{F}(10n_{f}+N_{C}(-67 + 3\pi^{2}))\\
\ea
and for the $(-)$ case we have
\ba
&&K_{1}^{-}(x)=\frac{1}{72}C_{F} (-2 n_{f} (3+4\pi^{2}) + N_{C}(51 + 44\pi^{2} - 216 \zeta(3))+ 9C_{F}(3-4\pi^{2}+48\zeta(3))\nonumber\\
&&K_{2}^{-}(x)= \frac{2 C_{F}(+2C_{F}-N_{C})x}{1+x}\\
&&K_{3}^{-}(x)= \frac{C_{F}(9C_{F}-11 N_{C}+2n_{f})x}{3(x-1)}\\
&&K_{4}^{-}(x)=\frac{C_{F}N_{C}x}{1-x}\\
&&K_{5}^{-}(x)=\frac{4C_{F}^{2}x}{1-x}\\
&&K_{6}^{-}(x)=-\frac{1}{9}C_{F}(10n_{f}+N_{C}(-67 + 3\pi^{2}))\\
&&K_{7}^{-}(x)=-\frac{1}{9}C_{F}(10n_{f}-18 C_{F}(x-1)+N_{C}(-76 +3\pi^{2}+9x))\\
\ea


\normalsize
\begin{thebibliography}{99}
\bibitem{Ji} X. Ji, Phys.Rev.D55 (1997) 7114.
\bibitem{Radyushkin}A.V. Radyushkin, Phys.Rev.D56 (1997) 5524.
\bibitem{Teryaev1} C. Bourrely, J. Soffer, and O.V. Teryaev Phys.Lett.B420 (1998) 375.
\bibitem{Teryaev} C. Bourrely, E. Leader, O.V. Teryaev,  hep-ph/9803238.
\bibitem{JJ} R.L. Jaffe and X. Ji, Phys. Rev. Lett. 67 (1991) 552.
\bibitem{CafaCor} A. Cafarella and C. Corian\`{o}, hep-ph/0301103.
\bibitem{Barone} E. Barone, Phys.Lett.B409 (1997) 499.
\bibitem{Gordon} L.E. Gordon and G. P. Ramsey, Phys.Rev.D59 (1999) 074018.
\bibitem{Coriano}C. Corian\`{o}, Nucl.Phys.B627:66,2002. 
\bibitem{CollinsQiu} J.C. Collins and J. Qiu, Phys.Rev.D39 (1989) 1398.
\bibitem{teryaevsoffer} C. Bourrely, J. Soffer, O.V. Teryaev, Phys.Lett.B420 (1998) 375.
\bibitem{GRV}M.Glück, E.Reya and A.Vogt, Eur.Phys.J.C 5 (1998) 461
\bibitem{GRSV}M.Glück et al., Phys.Rev.D 63 (2001) 094005
\bibitem{MSSV}O.Martin et al., Phys.Rev.D 57 (1998)
\bibitem{Rossi} G. Rossi, Phys.Rev.D29 (1984) 852.
\bibitem{Golec} K.J. Golec-Biernat and A. D. Martin, Phys.Rev.D59 (1999) 014029.
\bibitem{GRV}M.Glück, E.Reya and A.Vogt, Eur.Phys.J.C 5 (1998) 461
\bibitem{GRSV}M.Glück, E.Reya, M.Stratmann and W.Vogelsang, Phys.Rev.D 63 (2001) 094005
\bibitem{CTEQ}H.L.Lai \emph{et al.}, Phys.Rev.D 55 (1997) 1280.
\bibitem{GGR}L.E.Gordon, M.Goshtasbpour and G.P.Ramsey, Phys.Rev.D 58 (1998) 094017.
\bibitem{MSSV}O.Martin, A.Schäfer, M.Stratmann and W.Vogelsang, Phys.Rev.D 57 (1998) 
117502.
\bibitem{ERBL} A.V. Efremov and A.V. Radyushkin, Phys. Lett. B94 (1980) 245;
G. P. Lepage and S.J. Brodsky, Phys. Rev. D22 (1980) 2157;
\bibitem{CL} C. Corian\`{o} and H. N. Li, JHEP 9807 (1998) 008.
\end{thebibliography}
\end{document}